\documentclass[10pt,aps,prd,showpacs,superscriptaddress,preprintnumbers,nofootinbib]{revtex4-1}
\usepackage{graphicx,amsmath,amssymb,bbm,enumitem,hyperref,xspace,subfigure}
\usepackage[dvipsnames]{xcolor}
\newcommand{\be}{\begin{eqnarray}}
\newcommand{\ee}{\end{eqnarray}}

\newcommand{\barr}[1]{\bar{\bar{#1}}}
\newcommand{\ie}{i.e.\xspace}
\newcommand{\eg}{e.g.\xspace}
\newcommand{\twototwo}{$2\to2$\xspace}
\newcommand{\threetothree}{$3\to3$\xspace}
\newcommand{\perm}{\ensuremath{\Pi}\xspace}
\allowdisplaybreaks
 
\begin{document}

\title{Three-body Unitarity with Isobars Revisited}

\author{M. Mai}
\email{maximmai@gwu.edu}
\affiliation{The George Washington University, Washington, DC 20052, USA}

\author{B. Hu}
\affiliation{The George Washington University, Washington, DC 20052, USA}

\author{M.\ D\"oring}
\affiliation{The George Washington University, Washington, DC 20052, USA}
\affiliation{Thomas Jefferson National Accelerator Facility, Newport News, VA
23606, USA}

\author{A. Pilloni}
\affiliation{Thomas Jefferson National Accelerator Facility, Newport News, VA
23606, USA}

\author{A. Szczepaniak}
\affiliation{Thomas Jefferson National Accelerator Facility, Newport News, VA
23606, USA}
\affiliation{Center for Exploration of Energy and Matter, Indiana University, Bloomington, IN 47403, USA}
\affiliation{Physics Department, Indiana University, Bloomington, IN 47405, USA}

\preprint{JLAB-THY-17-2496}

\begin{abstract}
The particle exchange  model of hadron interactions can be used to describe three-body scattering under the isobar assumption. 
In this study we start from the \threetothree scattering amplitude for spinless particles, which contains an isobar-spectator scattering amplitude. Using a Bethe-Salpeter Ansatz for the latter, we derive a relativistic three-dimensional scattering equation that manifestly fulfills three-body unitarity and two-body unitarity for the sub-amplitudes.
This property holds for energies above breakup and also in the presence of resonances in the sub-amplitudes. 
\end{abstract}

\pacs{
11.80.−m, 
11.80.Cr, 
11.80.Et, 
11.80.Jy 
}
\maketitle

\section{Introduction}\label{sec:intro}
There is a growing interest in the description of 
three-hadron systems, as new phenomena that are being discovered  may critically depend on three-particle dynamics. For example, light hybrids and other meson resonances with mass above $1 \mbox{ GeV}$ are extensively studied at COMPASS with pion beams~\cite{Abbon:2007pq} and in the near future will be produced with photons at GlueX and CLAS12~\cite{ Ghoul:2015ifw, Glazier:2015cpa}. They are expected to have large branching ratios to three pions and, even tough there are no known resonances generated by three pion interactions,  there are features in the spectrum such as, \eg, the appearance of the $a_1(1420)$ enhancement,  that can be attributed to irreducible three-body interactions~\cite{Ketzer:2015tqa}. Similarly, one expects to reach a clearer picture of the $XYZ$ sector~\cite{Lebed:2016hpi, Esposito:2016noz, Guo:2017jvc} currently explored by LHCb, BESIII, Belle and BaBar~\cite{Alves:2008zz, Fang:2016gsh,pbf}, when the three-body interactions are taken into account. Three-body scattering plays also an important role in the pion-nucleon system~\cite{Aitchison:1978pw}, as is known from the large $\pi\pi N$ branching ratios of many resonances. In parallel there have been new  developments in establishing the formalism for the extraction of the three-particle scattering amplitudes in infinite volume from the energy eigenvalues from lattice QCD simulations~\cite{Hansen:2014eka, Hansen:2015zga, Briceno:2017tce, Meissner:2014dea, Polejaeva:2012ut, Roca:2012rx, Agadjanov:2016mao,Hammer:2017uqm,Hammer:2017kms} and the future analysis of lattice data will require knowledge of the analytical three-body scattering amplitudes.

Determination of the underlying dynamics from experimental and lattice data requires as input analytical reaction amplitudes. For scattering in the resonance region, unitarity plays the key role in constraining the analytical properties of $S$-matrix elements. 
In general, models of reaction amplitudes that focus on resonance physics can be classified as either non-linear on-energy-shell, or linear, off-energy-shell. The former are based directly on the $S$-matrix principles, that result, through dispersion  relations, in non-linear constraints on partial wave amplitudes. The latter derive from effective models of particle exchanges which are used as kernels of scattering  amplitudes. An example of an $S$-matrix constraint on an amplitude involving three-particle final states are the Khuri-Treiman equations~\cite{Khuri:1960zz,Niecknig:2012sj,Schneider:2012ez,Danilkin:2014cra,Guo:2015zqa,Niecknig:2015ija,Guo:2016wsi,Albaladejo:2017hhj,Isken:2017dkw}, that have been used extensively to describe three-particle decays. In these analyses, two-body unitarity is implemented in all the sub-channels, however, it should be noted that the mass of the decaying particle is treated as a mere parameter, and the dependence of the amplitude on it is unconstrained. To enforce that, one needs to consider the full \threetothree amplitude, where the decaying particle appears as a pole in the total invariant mass, and impose three-body unitarity. 

Implementing three-body unitarity in re-scattering schemes, such as provided by the Bethe-Salpeter equation, is much more intricate than in the two-body case. 
Different approaches are available to construct \threetothree amplitudes. The complexity of full Faddeev calculations is considerable, see \eg the AGS formulation~\cite{Alt:1967fx}. Three-body extensions of chiral two-body amplitudes in Faddeev-like manner~\cite{MartinezTorres:2007sr, Magalhaes:2011sh, MartinezTorres:2011vh} are usually restricted to quantum numbers in which all particles are in relative $S$-wave. This limits their usefulness as analysis tool. Another frequently used approach for the interaction of three particles is the fixed-center approximation for systems in which two interacting bodies are much heavier than the spectator. Recently, such an approach was applied to $\bar Kd$ scattering considering systematic corrections to this approximation~\cite{Mai:2014uma}.

In this paper we use the framework developed in~\cite{Aaron:1969my,Aaron:1973ca, Amado:1974za,Aaron:1975wc, Amado:1975zz, Aaron:1975iq,Aaron:1976cf, AAYBook}, which we refer to as the Amado model. The model provides an appealing solution to the problem, because it can be reduced to a three-dimensional integral equation through the isobar assumption. The definition of an isobar is given in the next section. The model has inspired many three-body formalisms, in particular through the concept of hadron exchange that arises as a necessary topology in the interaction. In particular, dynamical coupled-channel models~\cite{Ronchen:2012eg, Ronchen:2015vfa, Kamano:2013iva, Kamano:2014zba} formulate their interactions in terms of the exchanges of light mesons and baryons, together with the possibility of an in-flight decay of the isobar. Related approaches~\cite{Baru:2015nea, Baru:2015tfa, Baru:2011rs, Mai:2014uma, Mizutani:2012gy, Longacre:1990uc, Afnan:1988fk, Thomas:1976px, Avishai:1979nm, Mennessier:1972bi, Ceci:2011ae, Doring:2009yv,Kamano:2011ih,Nakamura:2015qga} also use the isobar-spectator picture for a variety of scattering problems.

Specifically, we focus the discussion on the Aaron, Amado and Young papers (AAY in the following)~\cite{Aaron:1969my,AAYBook} in which the amplitude is constructed by matching to unitarity. The original work claims that the derived amplitudes fulfill three-body unitarity, but the proof is only valid for the bound-state particle scattering. At energies above breakup, the proof does not hold. We are going to present a more careful derivation of matching equations, showing that the proof can be extended to any energies and isobars with both bound states and resonances. The paper is organized as follows. In Sec.~\ref{sec:isobar}, the \threetothree amplitude is formulated using the isobar assumption. The unitarity relation is discussed qualitatively in Sec.~\ref{sec:unitarity}, while explicit analytic expressions can be found in Appendix~\ref{sec:appendix_a}. In Sec.~\ref{sec:bse} the Bethe-Salpeter Ansatz is defined and expanded, such that it matches the right-hand side of the unitarity relation term by term. The matching, which allows to determine the building blocks of the Bethe-Salpeter equation, is performed in Sec.~\ref{sec:matching}. There, the method is also compared to AAY~\cite{Aaron:1969my}, and the parts missing in their derivation are identified.

\section{\texorpdfstring{The Isobar approximation and \threetothree scattering}{The Isobar approximation and 3->3 scattering} }\label{sec:isobar}

We consider the scattering of three identical, asymptotically stable, spinless, isoscalar particles $\cal{P}$ of mass $m$ (we refer to this simply as ``meson'' in the following). For \threetothree scattering, the isobar assumption states that the incoming ($|p_1,p_2,p_3\rangle$) and outgoing three-particle states ($\langle q_1,q_2,q_3|$) are only populated through a {\it spectator} $\cal{P}$ and an {\it isobar} $\cal{P}^*$ decaying into two particles, \ie, $\cal{P}[\cal{P}^*\to \cal{PP}]\cal{P}$. Therefore, the isobar-spectator propagation is a function of the total energy $\sqrt{s}$ and the (on-shell) momentum of the spectator. 
In this context, the isobar parameterizes the dynamics of a subchannel partial-wave. As a partial-wave, it has definite quantum numbers, \ie spin and parity,  and depends only on the invariant mass of the two particle subsystem.   It has complexity arising from direct channel unitarity, and possibly from exchange interactions at unphysical values of the subchannel energy. We note, however, that the dependence of the phase of the full \threetothree amplitude on the isobar mass is different from the one of the corresponding partial wave in elastic scattering, because of direct channel contributions from isobar-spectator interactions~\cite{Guo:2014vya}. The isobar approximation is expected to hold as long as the total energy of the three-particle system is low, i.e., when the physical $3\to 3$ amplitude can be approximated by a finite number of partial waves, and the subchannel amplitude is dominated by the right-hand singularities (either threshold branch points, or resonance/bound state poles).  At high energy, the details of the left hand singularities will dominate the amplitude, and the approximation breaks down. We remark that we do not need the isobar to be dominated by any resonance -- the same formalism can be applied, say, to a $\pi^+\pi^+$ repulsive subchannel.

\begin{figure}
\begin{center}
\includegraphics[width=1.\linewidth]{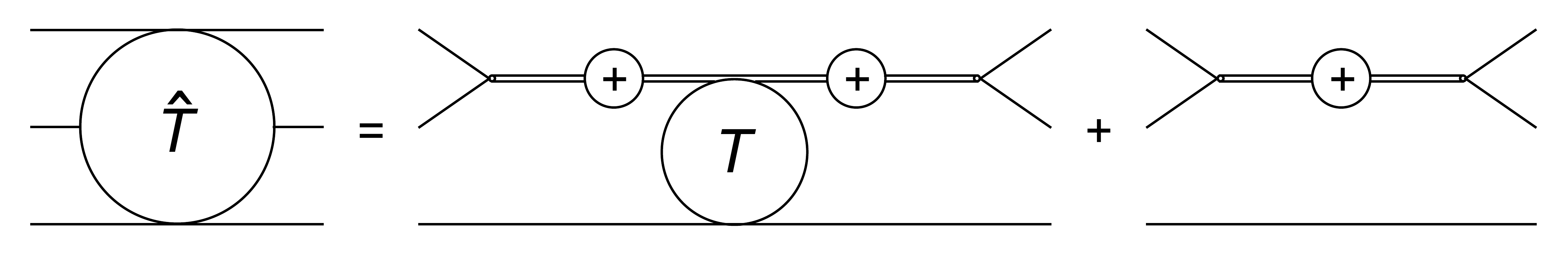}
\end{center}
\caption{Total scattering amplitude $\hat T$ consisting of a connected ($\hat T_c$) and a disconnected contribution ($\hat T_d$), represented by the first and second term on the right-hand side, respectively. Single lines indicate the elementary particle, double lines represent the isobar, and time runs from right to left in this and subsequent figures. In the isobar approximation three-body states can only be populated via an isobar-spectator state $S$ with subsequent isobar decay $\cal P^*\to \cal P\cal P$. The state $S$ is indicated as ``+'' and the (amputated) isobar-spectator scattering amplitude as $T$.}
\label{fig:T_total}
\end{figure}

In Fig.~\ref{fig:T_total}, the \threetothree scattering process is shown. The isobar propagator $S$ is symbolized with ``+'' and followed/preceded by isobar decay/formation via a decay function $\tilde v$. This may be a ``vertex'' in the sense that it is derived from some Lagrangian, or simply a function of the isobar invariant mass and angles that has no singularities related to two-particle interactions in the direct channel. The angular dependence is meant to give the isobar definite spin, but will be treated generically in the following.  The resulting scattering amplitude $\hat T$ consists of a connected ($\hat T_c$) and a disconnected piece ($\hat T_d$). In particular, there is a yet unknown isobar-spectator interaction $T$ that will be determined by matching a Bethe-Salpeter equation (BSE) to the unitarity relation in the following. We adopt here the convention that time flows from the right to left so that the diagram and operator representations of the amplitude appear in the same order.

Let us specify the numbering of external four-momenta, \eg, as $q_1$ for the spectator and $q_2$, $q_3$ for the two particles from the isobar decay. Then we write $S((q_2+q_3)^2)$ and $\tilde v(q_2,q_3)$, where the latter is considered to be symmetric under the exchange of both elementary particles.
In what follows we consider the case of only one isobar. The relaxation of this assumption is straightforward, \eg, by embedding the propagators of all considered isobars as elements of diagonal matrices. However, since the main goal of this work is the analysis of the implications of unitarity, we leave the generalization to future work. For the same reason, we neither specify the quantum numbers of the isobar, nor the exact form of the vertex $\tilde v$. In order to make the comparison to the AAY result~\cite{AAYBook,Aaron:1969my} more instructive, one can think of an $S$-wave isobar with a momentum-independent coupling to the asymptotic states, \ie $\tilde v(q_2,q_3)=\lambda/2$. We will perform such a comparison at a later stage of the present analysis, see Section~\ref{sec:final}.

The scattering amplitude $T$ for the isobar-spectator interaction ${\cal P}^*(P-p_m){\cal P}(p_m)\to {\cal P}^*(P-q_n){\cal P} (q_n)$  is given by $\langle q_n|T(s)|p_m\rangle:=\langle P-q_n,q_n|T(s)|P-p_m,p_m\rangle$ for the total four-momentum of the three-body system  $P$ so that $P^2=s$. With this, the fully connected part $\hat T_c$ of the scattering amplitude $\hat T$ in Fig.~\ref{fig:T_total} is written as
\begin{align}
\label{eq:tc0}
\langle q_1,q_2,q_3|\hat T_c(s)| &p_1,p_2,p_3\rangle=\\\nonumber
\frac{1}{3!}&\sum_{\perm\in S^3, \perm'\in S^3}
\tilde v(q_{\perm(2)},q_{\perm(3)}) S((q_{\perm(2)}+q_{\perm(3)})^2)%
\langle q_{\perm(1)}|T(s)|p_{\perm'(1)}\rangle
S((p_{\perm'(2)}+p_{\perm'(3)})^2)\tilde v(p_{\perm'(2)},p_{\perm'(3)}) \,,
\end{align}
where $p_m (q_n)$ are the in-(out-)going on-shell four-momenta, $p_m^2=q_n^2=m^2$. The expression on the right-hand side accounts for all possible assignments of momenta, where $\perm^{(}{}'{}^{)}$ are permutations of the sequence $\{1,2,3\}$, as well as for the overall normalization factor $1/3!$. To abbreviate the notation further, we re-define for any given spectator with particle index $n$ the dissociation vertex $v(q_{\bar n},q_{\barr{n}}):=2\tilde v(q_{\bar n},q_{\barr{n}})$, where $\bar n$ and $\bar{\bar n}$ are particle indices from $\{1,2,3\}$, which are neither equal among each other nor equal to $n$. It is exactly due to the symmetry of $\tilde v$ that the sums over permutations $\perm^{(}{}'{}^{)}$ simplify, becoming simple sums
\begin{align}
\label{eq:tc1}
\langle q_1,q_2,q_3|\hat T_c(s)| p_1,p_2,p_3\rangle
=\frac{1}{3!}\sum_{n=1}^3\sum_{m=1}^3
v(q_{\bar{n}},q_{\barr{n}}) S(\sigma(q_n))\,
\langle q_n|T(s)|p_m\rangle\,
S(\sigma(p_m))v(p_{\bar{m}},p_{\barr{m}}) \,,
\end{align}
where $\sigma(q):=(P-q)^2$ is the invariant mass of the isobar squared. Note that there are 36 terms on the right-hand side in Eq.~(\ref{eq:tc0}) but only 9 terms on the right-hand side of Eq.~\eqref{eq:tc1}. This is a consequence of the discussed redefinition $\tilde v\to v$ and the symmetry of the vertex function under exchange of identical particles. In Eq.~(\ref{eq:tc1}) the sums run over the incoming (index $m$) and outgoing spectator momenta (index $n$).

Fig.~\ref{fig:T_total} contains also a disconnected part of the \threetothree scattering which is given by 
\begin{align}
\label{eq:tdisc}
\langle q_1,q_2,q_3|\hat T_d(s)| p_1,p_2,p_3\rangle
&=
\frac{1}{3!}\sum_{n=1}^3\sum_{m=1}^3\,
\langle q_n|\mathbbm{1}|p_m\rangle
v(q_{\bar{n}},q_{\barr{n}}) S(\sigma(q_n)) v(p_{\bar{m}},p_{\barr{m}})
\\ \nonumber
&=
\frac{1}{3!}\sum_{n=1}^3\sum_{m=1}^3\,
2E_{q_n}(2\pi)^3\delta^3(\mathbf{q}_n-\mathbf{p}_m)\,
v(q_{\bar{n}},q_{\barr{n}}) S(\sigma(q_n)) v(p_{\bar{m}},p_{\barr{m}}) \,.
\end{align}
Note that the external particles are always on their mass shell which means that the identity operator $\mathbbm{1}$ acts on the bra/ket states each of which being a function of corresponding three-momenta only. This is the usual convention, see, \eg, chapter 4.6 of \cite{Peskin}, which however differs from the one chosen in \cite{Aaron:1969my}. Both choices are legitimate. However, in the following we will stick to the convention of Ref.~\cite{Peskin} and thus the same one has to be chosen for the above normalization condition, \ie $\langle q_n|p_m\rangle=2E_{q_n}(2\pi)^3\delta^3(\mathbf{q}_n-\mathbf{p}_m)$ where $E_q=\sqrt{{\bf q}^2+m^2}$.

The full \threetothree amplitude is given by
\begin{align}
\label{eq:t33full}
\langle q_1,q_2,q_3|\hat T(s)| p_1,p_2,p_3\rangle
=
\langle q_1,q_2,q_3|\hat T_c(s) |p_1,p_2,p_3\rangle+\langle q_1,q_2,q_3|\hat T_d(s)| p_1,p_2,p_3\rangle\,.
\end{align}
As indicated before the external particles are on the mass shell. Moreover, due to energy-momentum conservation, the number of free variables of a \threetothree scattering amplitude is 8, which, \eg, in the center-of-mass frame of the three particles can be chosen to be the total energy squared (1), the angle between the incoming and outgoing spectator (1) in the scattering plane, invariant mass of the incoming and outgoing isobars (2) as well as the angle of the isobar decay plane with the scattering plane (2) and the angle between the isobar and one of its decay products (2).

\section{Three-Body Unitarity }\label{sec:unitarity}

\begin{figure*}
\begin{center}
\includegraphics[width=1.\linewidth,trim=1cm 3.5cm 1cm 0.8cm]{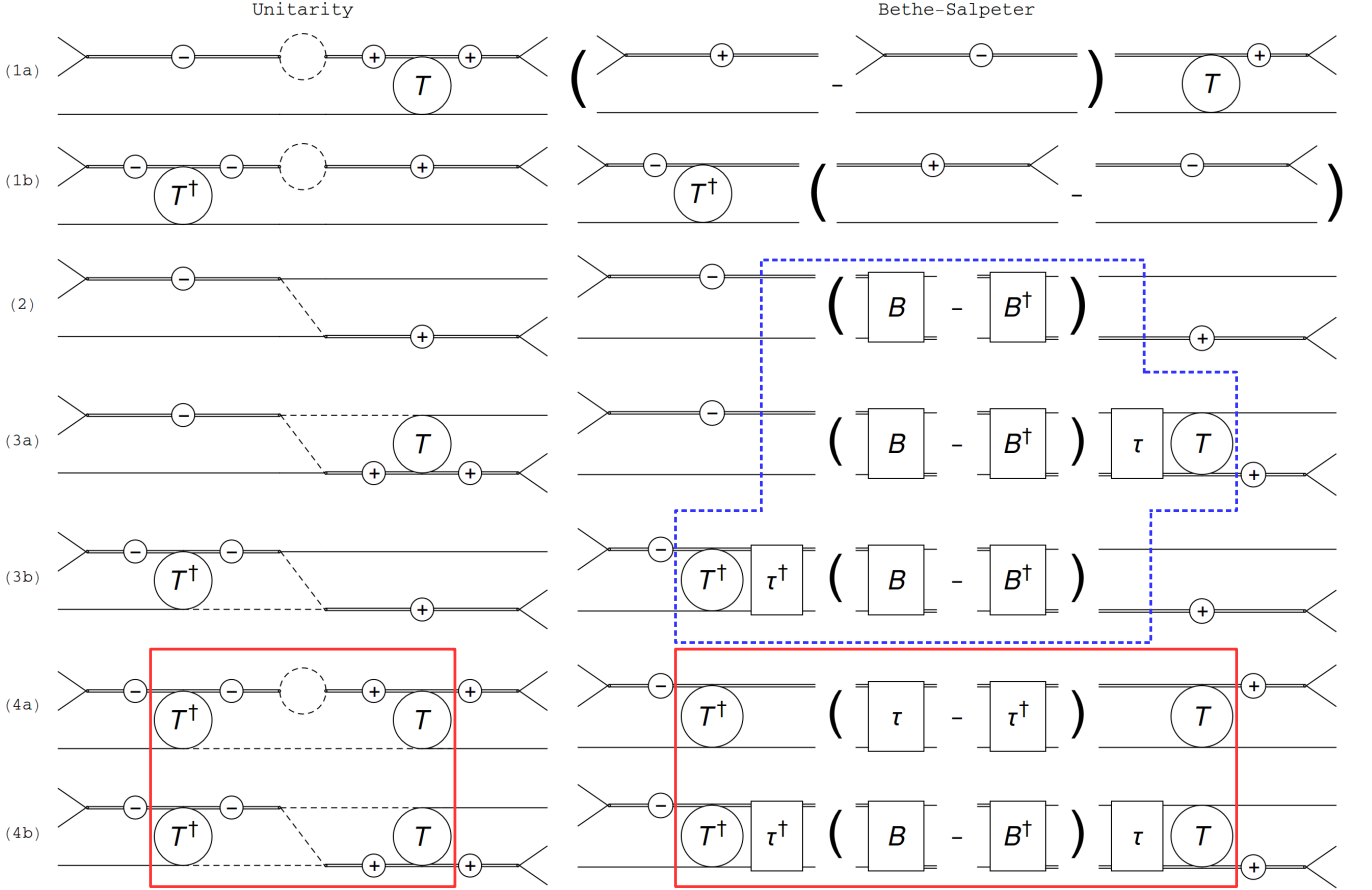}
\end{center}
\caption{Left: Graphical representation of the right-hand side of the unitarity relation (\ref{eq:unit0}). Right: Expansion of the Bethe-Salpeter amplitude~\eqref{eq:bse1} according to Eqs.~(\ref{eq:tstep}, \ref{eq:bse3}).
See Fig.~\ref{fig:T_total} and text for further notation. The black dashed lines indicate intermediate states at momentum $k_\ell$ that are set on-shell by the $\delta^+$-distribution in Eq.~\eqref{eq:unit0}. The parts inside the red solid boxes represent the sub-processes considered by AAY~\cite{Aaron:1969my} for the  matching. The parts inside the blue dashed outline contain terms that unavoidably arise in the expansion of the BSE (see Eq.~(\ref{eq:bse3})) and that only vanish for bound-state scattering and below breakup, $\sqrt{s}<3m$. They have no counterpart in the unitarity relation in the formulation of Ref.~\cite{Aaron:1969my}.  
}
\label{fig:compare}
\end{figure*}

The unitarity of the $\cal{S}$-matrix ($\mathcal{S}{}:=\mathbbm{1}+i(2\pi)^4 \delta^4\!\left(P_{i}-P_{f}\right) \hat T$) leads to the following condition for the scattering amplitude
\begin{align}
\label{eq:unit0}
\langle q_1,q_2,q_3|&(\hat T-\hat T^\dagger)| p_1,p_2,p_3\rangle=\\
&i\int\prod_{\ell=1}^3\left[\frac{\mathrm{d}^4k_\ell}{(2\pi)^{4}}\,(2\pi)\delta^+(k_\ell^2-m^2)\right]
\,(2\pi)^4\delta^4\left(P-\sum_{\ell=1}^3\,k_\ell\right)\langle q_1,q_2,q_3|\hat T^\dagger|k_1,k_2,k_3\rangle\,
\langle k_1,k_2,k_3|\hat T| p_1,p_2,p_3\rangle \,,\nonumber
\end{align}
where we have inserted a complete set of states in the convention of Ref.~\cite{Peskin}, also utilized in Ref.~\cite{AAYBook}. Specifically, the unitarity relation does not include the factor $1/3!$, which is already accounted for in the definition of the scattering amplitude, see Eqs.~\eqref{eq:tc1} and \eqref{eq:tdisc}. We emphasize that the unitarity condition is formulated in terms of on-shell to on-shell transitions for asymptotically stable particles. In contrast, the proof of Ref.~\cite{Aaron:1969my} considers only the isobar-spectator scattering process $T$ of Eq.~(\ref{eq:tc0}). 
In Eq.~(\ref{eq:unit0}), the function $\delta^+(k_\ell^2-m^2)=\delta(k_\ell^2-m^2)\theta(k_\ell^0)$ forces the intermediate states at momenta $k_\ell$ to be on-shell with positive energy.

Expanding the right-hand side of Eq.~(\ref{eq:unit0}) using Eqs.~(\ref{eq:tc1}, \ref{eq:tdisc}) one obtains seven connected and one disconnected topologies. The latter addresses two-body unitarity only and, thus, does not lead to any new insights regarding the \threetothree scattering amplitude. The fully connected
topologies are shown in Fig.~\ref{fig:compare} to the left. For the isobar propagation one obtains factors of $S$, indicated with ``+'' as before, but also of the form $S^\dagger$, indicated with the ``$-$'' sign. The dashed lines indicate intermediate states at momentum $k_\ell$ that are set on-shell by the $\delta^+$-distribution in Eq.~\eqref{eq:unit0}.

The combination $\hat T_d^\dagger\hat T_d$ in Eq.~(\ref{eq:unit0}) can also lead to a fully connected topology shown as diagram (2) to the left in Fig.~\ref{fig:compare}. Disconnected-connected combinations of $\hat T_d^\dagger\hat T_c$ and $\hat T_c^\dagger\hat T_d$ lead to the topologies (1) and (3) in Fig.~\ref{fig:compare} while connected-connected contributions lead to the diagrams (4). To illustrate the origin of the different topologies we consider diagram (1a) and (3a) arising from the combination $\hat T_d^\dagger\hat T_c$ in the unitarity relation. Considering (for illustration) only this combination, denoted by index ``$dc$'', Eq.~(\ref{eq:unit0}) reads 
\begin{align} 
\label{eq:unit1a}
\langle q_1,q_2,q_3|(\hat T-\hat T^\dagger)| p_1,p_2,p_3\rangle_{dc}\nonumber
=&i\int\prod_{\ell=1}^3\left[\frac{\mathrm{d}^4k_\ell}{(2\pi)^{4}}\,(2\pi)\delta^+(k_\ell^2-m^2)\right]\,
(2\pi)^4\delta^4\left(P-\sum_{\ell=1}^3\,k_\ell\right)\\\nonumber
&\times\frac{1}{3!}\,\sum_{n,i}
v(q_{\bar{n}},q_{\barr{n}})
S^\dagger (\sigma(q_n))v(k_{\bar{i}},k_{\barr{i}})
\,
2E_{\mathbf{q}_n}(2\pi)^3\delta^3(\mathbf{q}_n-\mathbf{k}_i) \\\nonumber
&\times
\frac{1}{3!}\,\sum_{j,m} v(k_{\bar{j}},k_{\barr{j}})S(\sigma(k_j))\langle k_j|T|p_m\rangle S(\sigma(p_m))v(p_{\bar{m}},p_{\barr{m}})\\
&\times \big[\underbrace{\delta_{ij}}_{\to\,\text{(1a)}}+\underbrace{(1-\delta_{ij})}_{\to\,\text{(3a)}}\big]\,.\hspace{10cm}
\end{align}
The last line shows that in the sums over internal states the momentum labels can be arranged such that the spectators are the same $(i=j)$ or different ($i\neq j$), leading to the topologies (1a) and (3a) shown to the left in Fig.~\ref{fig:compare}, respectively. The expressions corresponding to the other topologies are quoted in Appendix~\ref{sec:appendix_a}.

In the original formulation by AAY~\cite{Aaron:1969my} the unitarity constraint is applied on the isobar-spectator interaction $T$ directly.
Therefore, only the topologies (4a) and (4b) are present there, without the external factors $S\,v$, see Eq.~(\ref{eq:tc1}) or Fig.~\ref{fig:T_total}. These parts are framed by the red solid lines in Fig.~\ref{fig:compare}. This oversimplification is the origin of the incompleteness of the argumentation of Ref.~\cite{Aaron:1969my} as discussed in the following sections.

\section{Bethe-Salpeter Ansatz }\label{sec:bse}

\begin{figure*}
\begin{center}
\includegraphics[width=1.\linewidth]{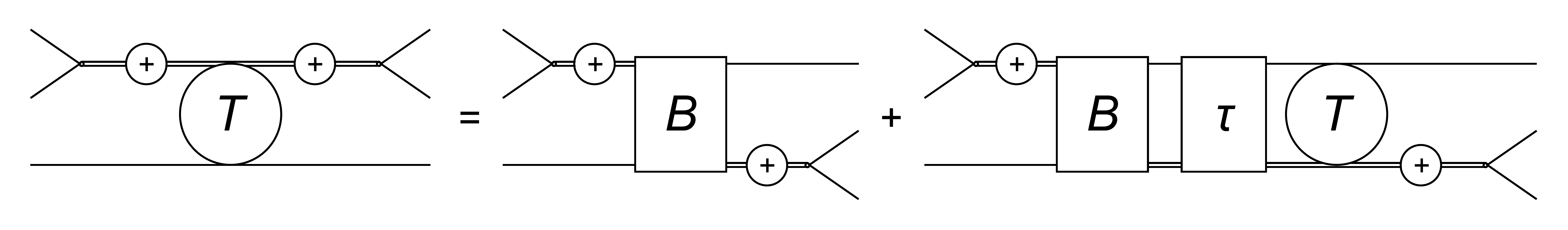}
\end{center}
\caption{Bethe-Salpeter Ansatz embedded into the \threetothree amplitude. 
$B$ denotes the yet unspecified driving term of the scattering amplitude, while $\tau$ is the Greeen's function corresponding to the propagation of the isobar and elementary particle.}
\label{fig:BSE}
\end{figure*}

The connected part of the \threetothree scattering amplitude contains a yet unknown piece, namely the isobar-spectator scattering amplitude $T$. To access this piece we re-write the discontinuity in operator notation\footnote{In this work we make no difference in notation between operators and amplitudes as misunderstandings should be excluded.},
\begin{align}\label{eq:tstep}
\hat T-\hat T^\dagger&=
v(S-S^\dagger) v
+
vSTSv-vS^\dagger T^{\dagger}S^\dagger v\nonumber \\
&=
v(S-S^\dagger) v
+
\underbrace{\left(vS-vS^\dagger \right)TSv}_{\to\text{(1a)}}
+\underbrace{vS^\dagger T^{\dagger}\left(Sv-S^\dagger v\right)}_{\to\text{(1b)}}
+vS^\dagger (T-T^{\dagger})Sv \,.
\end{align}
The term $v(S-S^\dagger)v$ is the fully disconnected contribution to the right-hand side of the unitarity relation and will not be considered further for the reasons described in the previous section. The second and third terms indicate structures of the topologies (1a) and (1b) as derived from the unitarity relation. We symbolize them in the second column of Fig.~\ref{fig:compare}. The last term on the right-hand side of the above equation $vS^\dagger(T-T^\dagger)Sv$ cannot be further evaluated without making an Ansatz for the isobar-particle interaction $T$. We proceed along the steps of Ref.~\cite{Aaron:1969my}, making a Bethe-Salpeter equation (BSE) Ansatz for the $T$-matrix describing the ${\cal P}^*{\cal P}\to {\cal P}^*{\cal P}$ scattering. For any given in/outgoing spectator momenta $p_m/q_n$ (not necessarily on-shell), the BSE reads
\begin{align}
\label{eq:bse1}
\langle q_n|T(s)|p_m\rangle
=
\langle q_n|B(s)|p_m\rangle
+
\int \frac{\mathrm{d}^4 k}{(2\pi)^4}
\langle q_n|B(s)|k\rangle	\,
\tau(\sigma(k))			\,
\langle k|T(s)|p_m\rangle	\,.
\end{align}
Here, $\tau(\sigma(k))$ and $\langle q_n|B(s)|p_m\rangle$ denote the yet unknown isobar-spectator Green's function and the interaction kernel, respectively. In operator language the above Ansatz reads $T=B+B\tau T=B+T\tau B$ which is depicted symbolically in Fig.~\ref{fig:BSE}. As demonstrated in App.~\ref{sec:appendix_b} this Ansatz leads to
\begin{align}
\label{eq:bse3}
T-T^{\dagger}
=\underbrace{(B-B^\dagger)}_{\to\text{(2)}}
+\underbrace{(B-B^\dagger)\tau T}_{\to\text{(3a)}}
+\underbrace{T^{\dagger}\tau^{\dagger}(B-B^\dagger)}_{\to\text{(3b)}} 
+\underbrace{T^{\dagger}(\tau -\tau^{\dagger})T}_{\to\text{(4a)}}
+\underbrace{T^{\dagger}\tau^{\dagger}(B-B^\dagger)\tau T}_{\to\text{(4b)}} \,,
\end{align}
where the underbraces specify the corresponding contributions discussed in the previous section. Pictorially, these contributions are represented in Fig.~\ref{fig:compare} to the right. As desired,  there is a term-by-term correspondence between the right-hand side of the unitarity relation and the expansion of the BSE. As the figure suggests we can match $(B-B^\dagger)$ with the exchange processes shown to the left in Fig.~\ref{fig:compare} and, similarly, $(\tau -\tau^{\dagger})$ with the product of two propagators $S$ (see case (4a) in the figure). The remaining task is, thus, 1) to provide the matching relations; 2) to show that these matching relations are consistent, \eg, that they are the same for all topologies up to different kinematics, and finally 3) to determine the missing pieces of the \threetothree scattering amplitude, \ie $\tau(\sigma(k))$, $S(\sigma(k))$ and 
$\langle q_n|B(s)|p_m\rangle$ from the emerging matching relations.

\section{Matching to Unitarity Relation and Critical Discussion}\label{sec:matching}

We have identified all topologies of the discontinuity of the fully connected part of the \threetothree amplitude both from the unitarity constraint~\eqref{eq:unit0} as well as from the BSE Ansatz~\eqref{eq:bse1}. Now, we confront these two formulations to determine the yet unknown quantities $B$, $\tau$ and $S$. This comparison is depicted in Fig~\ref{fig:compare}, which already suggests the form of the desired quantities. To demonstrate how this comparison can be performed on a quantitative level, we start comparing topology (1a), see Eqs.~\eqref{eq:unit1a} and \eqref{eq:tstep}. Evaluating the $\delta_{ij}$ term of Eq.~(\ref{eq:unit1a}) as shown explicitly in Appendix~\ref{sec:appendix_a} and setting it equal to the corresponding term in Eq.~(\ref{eq:tstep}) one obtains the matching relation as given in Eq.~\eqref{eq:result1a}. Topologies (1b) and (4a) lead to the same kind of discontinuity relation for the isobar propagator if we identify ${\tau(\sigma(k)):= (2\pi)\delta^+(k^2-m^2)S(\sigma(k))}$. However, the discontinuity relation from the comparison of topology (4a) leads to most general case with respect to the validity range of the spectator four-momentum $k$. Specifically, $k$ can take all values with the restriction that $k^2=m^2$. It reads   
\begin{align}\label{eq:match_s}
S(\sigma(k))-&S^\dagger (\sigma(k))=
\,i\,\frac{S(\sigma(k))S^\dagger (\sigma(k))}{2(2\pi)^2}\\\nonumber
&\times\int \mathrm{d}^4\bar K\,
\delta^+\left(\left(\frac{P-k}{2}+\bar K  \right)^2-m^2\right)
\delta^+\left(\left(\frac{P-k}{2}-\bar K  \right)^2-m^2\right)
\left(v\left(\frac{P-k}{2}+\bar K,\frac{P-k}{2}-\bar K\right)\right)^2\,,
\end{align}
where the integration is performed over the loop momentum with two closed lines in Fig.~\ref{fig:compare}, (1a, 1b, 4a), to the left. Furthermore, one recognizes immediately that the above expression is consistent with two-body unitarity for the \twototwo scattering amplitude of the form $T_{22}=vSv$ in operator notation.

The remaining topologies of types (2, 3a, 3b and 4b) lead to the discontinuity relations of the driving term $B$ of the BSE presented in detail in Appendix~\ref{sec:appendix_a}. They are consistent with each other when making the same identification as before, \ie,
${\tau(\sigma(k)) = (2\pi)\delta^+(k^2-m^2)S(\sigma(k))}$. In this case the result of the matching of the type (4b) leads to the most general form of the discontinuity relation with respect to the range of validity in the four-momenta $p$, $q$ of the in- and outgoing spectators. These momenta are free except for the fact that $p^2=q^2=m^2$. The final and most general result for the discontinuity of $B$ reads
\begin{align}\label{eq:match_b}
\langle q|B|p \rangle - \langle q|B^\dagger|p \rangle =  
iv(P-p-q,q)(2\pi)\delta^+((P-q-p)^2-m^2)v(P-p-q,p) \,.
\end{align}
In the next section, the final form of the scattering equation using both matching relations (\ref{eq:match_s}) and (\ref{eq:match_b}) will be determined.

Before proceeding, we comment on the original derivation of matching relations in AAY~\cite{Aaron:1969my} compared to the present one. As pointed out before, AAY directly apply the unitarity relation to the isobar-spectator scattering amplitude $T$. As a consequence, the right-hand side of the unitarity relation is then given by the parts of the topologies (4a) and (4b) to the left in Fig.~\ref{fig:compare} highlighted by the red solid line. This corresponds to removing $vS^\dagger $ on the left and $Sv$ on the right from the full expression. Subsequently, AAY expand $T$ based on the same BSE Ansatz (\ref{eq:bse1}) leading to Eq.~(\ref{eq:bse3}). They then choose the terms indicated as (4a), (4b) in Eq.~(\ref{eq:bse3}) as highlighted by the red lines on the right-hand side of Fig.~\ref{fig:compare}, \ie, again only the isobar-spectator re-scattering parts without the coupling to the external (asymptotically stable) states via an isobar propagation and subsequent decay. They then compare to the ``amputated'' unitary condition highlighted in red to left in the figure.
As a result, AAY obtain the same matching relations as the ones deduced here, \ie, Eqs.~(\ref{eq:match_s}) and (\ref{eq:match_b}). 

While the result coincides formally, there are the following deficiencies in AAY's procedure: 
$1)$ For the matching AAY still had to discard the contributions (2), (3a), and (3b) from the expansion of the BSE of Eq.~(\ref{eq:bse3}), highlighted with the blue dashed lines in Fig.~\ref{fig:compare}. However, these terms contribute unavoidably even in the amputated formalism of AAY. In Ref.~\cite{AAYBook}, AAY discuss these contributions and argue that they disappear for bound-state particle scattering, which is a valid argument only for $\sqrt{s}<3m$. Through this deficiency, their proof is indeed restricted to bound-state particle scattering below breakup. In this manuscript, we have extended the proof to energies above breakup and for arbitrary isobar-spectator scattering with a resonant or non-resonant (even repulsive) isobar\footnote{As a side remark, we have not explicitly considered the presence of a bound state here, but it can be directly incorporated in our formalism following AAY.}. $2)$ Ultimately, as demonstrated here, the problem of the AAY paper is that they considered only the amputated isobar-spectator scattering $T$ instead of the full \threetothree amplitude $\hat T$. Indeed, the concept of unitarity is formulated in terms of asymptotically stable states and should be used in that sense only. If consequently formulated, both in the unitarity relation and in the Ansatz for the BSE, the emerging topologies can be compared term-by-term resulting in a set of consistent matching relations.


\section{Determination Of The Scattering Equation}\label{sec:final}

In the previous section we have derived the discontinuity relations for the driving term of the BSE Ansatz~\eqref{eq:bse1} and the isobar propagator. This knowledge can be utilized to derive their functional forms that  enter the BSE Ansatz~\eqref{eq:bse1} and ultimately complete the missing pieces of the \threetothree scattering amplitude Eq.~\eqref{eq:t33full}.

First, the discontinuity of the inverse of the isobar propagator $D(\sigma(k)):=-S^{-1}(\sigma (k))$ from Eq.~\eqref{eq:match_s} can be evaluated very conveniently in the reference frame of the isobar, \ie $\bar{P}:=(P-k)=(\sqrt{\sigma(k)},\mathbf{0})$ where $k$ is the momentum of the spectator. In this frame the integral part of the Eq.~\eqref{eq:match_s} reads
\begin{align}
{\rm Disc\,} D(\sigma(k))=&
\frac{i}{8\pi^2}
\int d^4\bar K\,
\delta^+\left(\left(\bar{P}/2+\bar K  \right)^2-m^2\right)
\delta^+\left(\left(\bar{P}/2-\bar K  \right)^2-m^2\right)
\Big(v\Big(\bar{P}/2+\bar K,\bar{P}/2-\bar K\Big)\Big)^2 \nonumber\\
=&
\frac{i}{64\pi^2}
\frac{1}{K_{{\rm cm}}}
\int \mathrm{d^3}\mathbf{\bar K}\,
\frac{\delta\left(|\mathbf{\bar K}|-K_{{\rm cm}}\right)}{\sqrt{(\mathbf{\bar K})^2+m^2}}
\left(
v\left(
      \begin{pmatrix}
      \sqrt{\mathbf{\bar{K}}^2+m^2}\\
      \mathbf{\bar K} 
      \end{pmatrix},
      \begin{pmatrix}
      \sqrt{\sigma(k)}-\sqrt{\mathbf{\bar{K}}^2+m^2}\\
      -\mathbf{\bar K}
      \end{pmatrix}
\right)\right)^2\,, \label{eq:discD}
\end{align}
where $K_{{\rm cm}}=\sqrt{\sigma(k)/4-m^2}$. In the course of this work we have been sloppy with the quantum numbers of the isobar, such as angular momentum or helicity indices. In principle these must be attached already at the level of the \threetothree Ansatz~\eqref{eq:t33full}, but also when introducing the BSE Ansatz in Eq.~\eqref{eq:bse1}. However, in this work we are concerned with the implications on the \threetothree scattering amplitude from unitarity only. For this we have kept the four-momentum dependence of the isobar decay vertex throughout the manuscript explicitly. To make a simple and more transparent connection to the analysis of Ref.~\cite{Aaron:1969my} let us again consider the case of one $S$-wave isobar, with $v(p,q)=\lambda$. In that case Eq.~\eqref{eq:discD} simplifies to
\begin{align}
\label{eq:discDaay}
{\rm Disc\,} D(\sigma(k))=
&
\frac{i}{8\pi}
\frac{K_{{\rm cm}}}{\sqrt{\sigma(k)}}\lambda^2 \, .
\end{align}
which allows to reconstruct $D(\sigma(k))$, \eg, by using a twice subtracted dispersion relation in $\sigma$,  
\begin{align}
\label{eq:DD}
D(\sigma(k))=
&
A+B\,\sigma(k)+\frac{\sigma(k)^2}{\pi}\int\limits_{4m^2}^{\infty}
d\sigma'\,\frac{{\rm Im}\,D(\sigma')}{\sigma'^2(\sigma'-\sigma(k)-i\epsilon)} \,,
\end{align}
which is a finite expression which, upon a suitable choice of the constants $A$ and $B$, can be identified as the dressed inverse propagator of the isobar.
\footnote{To see a connection to the result of AAY, one can derive the same result writing the one-loop contribution to the self-energy
\begin{equation*}
D(\sigma(k))=
\sigma(k)-M_0^2-\frac{1}{(2\pi)^3}\int d^3\boldsymbol{\ell} \frac{\lambda^2}{2E_{\ell}(\sigma(k)-4E_{\ell}^2+i\epsilon)}\,.
\end{equation*}
This integral exhibits a logarithmic divergence, which  cancels out with the one in $M_0^2$. The
latter plays a role of a bare mass in EFT language. In AAY, in addition, form factors are introduced in the 
vertex function, to make each term in $D(\sigma(k))$ finite.}

Second, the discontinuity of the driving term of the BSE Ansatz $B$ of Eq.~\eqref{eq:match_b} can be re-written in terms of the momentum transfer $Q := (P-p-q)$ and its squared value $u:=(P-p-q)^2$, where $P$, $p$ and $q$ denote the total four-momentum of the system, of the incoming and of the outgoing spectator, respectively, 

\begin{align}
{\rm Disc\,} B(u) 
:=\langle q|B|p\rangle-\langle q|B^\dagger|p \rangle
&= 2\pi i\,
\delta^+\left(u-m^2\right)
v\left(P-p-q,q\right)
v\left(P-p-q,p\right)\nonumber\\
\label{eq:discBkk}
&=2\pi i\,
\frac{\delta\left(E_Q - \sqrt{m^2 + \mathbf{Q}^2}\right)}{2 \sqrt{m^2 + \mathbf{Q}^2}}
v\left(Q,q\right)
v\left(Q,p\right) .
\end{align}
As stated in the beginning of the paper, we assume that the dissociation vertex has no non-analyticities in the physical region. Therefore, we can simply utilize an unsubtracted dispersion relation to obtain
\begin{align}
\langle q|B(s)|p\rangle 
=-\frac{v(Q,q)v(Q,p)}{2\sqrt{m^2 + \mathbf{Q}^2} \left(E_Q - \sqrt{m^2 + \mathbf{Q}^2}+i\epsilon \right)}  \,.
\end{align}

\begin{figure}
\begin{center}
\subfigure[~]{\includegraphics[width=.35\textwidth]{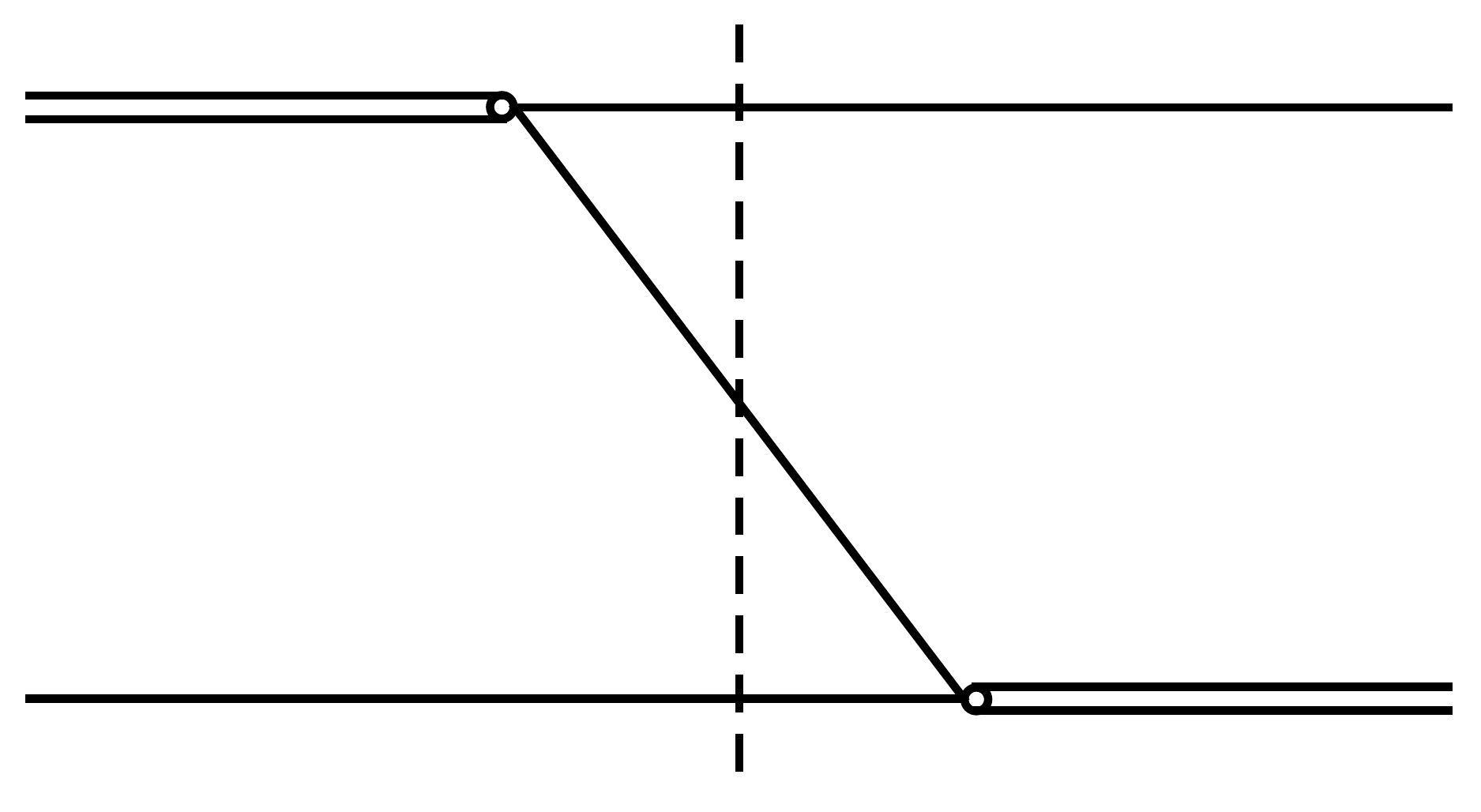}}
\hspace{1cm}
\subfigure[~]{\includegraphics[width=.50\textwidth]{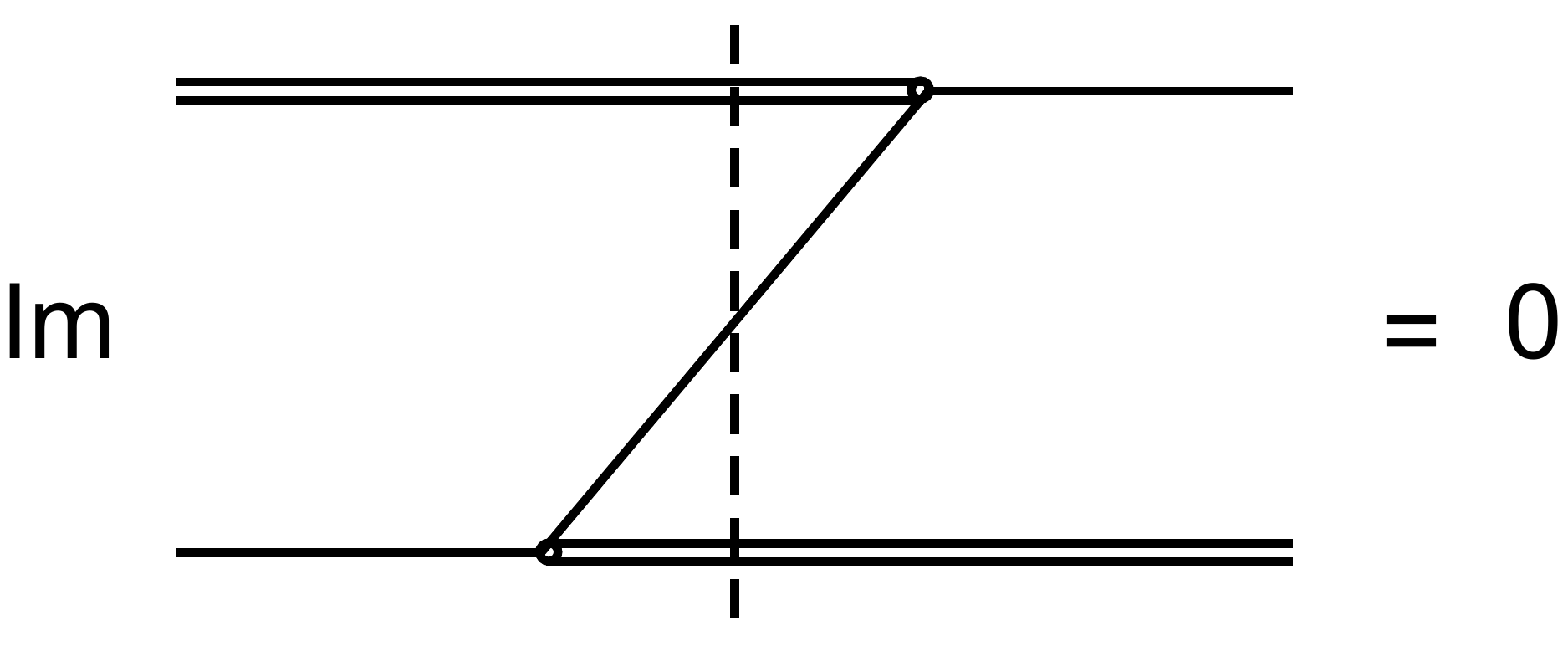}}
\end{center}
\caption{One-meson exchange contribution from Eq.~\eqref{eq:B}. Figure (a) shows the only contribution arising from unitarity, as expected in time-ordered perturbation theory. Figure (b) shows the additional diagram expected in time-ordered perturbation theory. It is real in the physical region, and can be added without altering Eqs.~(\ref{eq:match_b},\ref{eq:discBkk}), so as to restore covariance. }
\label{fig:timeordered}
\end{figure}

This terms resembles the diagram (a) in Fig.~\ref{fig:timeordered}, as expected in time-ordered perturbation theory.  This formal result is a possible solution which fulfills the discontinuity relations~(\ref{eq:match_b}, \ref{eq:discBkk}). We could, for example, add any real analytic function to the obtained result without altering Eqs.~(\ref{eq:match_b},\ref{eq:discBkk}). In particular, we are going to add the contribution (b) in Fig.~\ref{fig:timeordered}, which is real in the physical region, to restore covariance:  
\begin{align}
\langle q|B(s)|p\rangle 
&=-
\frac{v\left(Q,q\right)
v\left(Q,p\right)}{2 \sqrt{m^2 + \mathbf{Q}^2}\left(E_Q -\sqrt{m^2 + \mathbf{Q}^2} +i\epsilon\right)}
+\frac{v\left(Q,q\right)
v\left(Q,p\right)}{2 \sqrt{m^2 + \mathbf{Q}^2}\left(E_Q +\sqrt{m^2 + \mathbf{Q}^2}- i\epsilon\right)}\nonumber
 \\
&=\frac{v(P-p-q,q)v(P-p-q,p)}{m^2-u-i\epsilon} \label{eq:B} \, ,
\end{align}
where, in the last line, the $i\epsilon$ prescription is restored. This form is useful as it actually resembles the covariant form of a one-meson-exchange contribution. 

In summary, after the form for the \threetothree amplitude was chosen in Eqs.~\eqref{eq:tc1} and \eqref{eq:tdisc}, three-body unitarity has dictated the functional properties of the building blocks of the BSE. Note also that AAY derive a different result for the driving term of the BSE~\cite{Aaron:1969my, AAYBook} because the imaginary part~(\ref{eq:discBkk}) was dispersed in $s$ instead of $u$ as done here. Dispersing in $s$ implies that any other than the  $s$-dependence of this term is analytical and thus does not contribute to the discontinuity. Finally, AAY~\cite{Aaron:1969my, AAYBook} derive a Blankenbecler-Sugar type of particle exchange~\cite{Blankenbecler:1965gx}, which  does not exhibit the correct analytic behavior in the $u$-channel.

When we substitute the above equations into the Bethe-Salpeter Ansatz Eq.~\eqref{eq:bse1} we obtain our final expression after carrying out the integration over the zero momentum -- a three-dimensional integral equation that reads
\begin{align}\label{eq:final}
\langle q|T(s)|p\rangle =
& 
\frac{v(P-p-q,p)v(P-p-q,q)}{m^2-(P-p-q)^2-i\epsilon}
-
\int
\frac{\mathrm{d}^3\boldsymbol{\ell}}{(2\pi)^3}
\frac{1}{2E_\ell}
\frac{v(P-q-\ell,\ell)v(P-q-\ell,q)}{m^2-(P-q-\ell)^2-i\epsilon}
\frac{1}{D(\sigma(\ell))}
\langle \ell|T(s)|p\rangle\,,
\end{align}
where all momenta are meant to be on-energy-shell, \ie $E_\ell^2=(\boldsymbol{\ell}^2+m^2)$, $p_0^2=(\mathbf{p}^2+m^2)$ and ${q}_0^2=(\mathbf{q}^2+m^2)$. Furthermore, as defined before, $s=P^2$, with $P$ denoting the overall four-momentum of the isobar-spectator system.
This result differs from the one of AAY~\cite{Aaron:1969my} by the discussed form of $B$ and by the sign of the re-scattering term which comes from a re-definition $T\to- T$ made in Ref.~\cite{Aaron:1969my}.
Note that in Eq.~(\ref{eq:final}) $\sigma(\ell)=s+m^2-2\sqrt{s}E_\ell$, which implies $-\infty<\sigma(\ell)<\left(\sqrt{s} - m\right)^2$ in the integration domain.

In the simplest case of one $S$-wave isobar ($v(p,q)=\lambda$), the derivation of the \threetothree scattering amplitude is complete by substituting $T$ from Eq.~(\ref{eq:final}) into Eq.~\eqref{eq:t33full}. The explicit inclusion of higher spins and more than one isobar will be discussed elsewhere, see also discussion in the Section~\ref{sec:isobar}. Another interesting question is what implications arise in this framework for (infinitely) narrow resonances. As it is shown explicitly in Appendix~\ref{app:limit}, one obtains a unitary two-body scattering equation in the limit $\lambda\to 0$.

It is sometimes advantageous to formulate the \threetothree scattering amplitude in terms of the on-shell \twototwo scattering amplitude $T_{22}$. In the simplest case of one $S$-wave isobar and $v(p,q)=\lambda$, it reads $T_{22}(\sigma)=\lambda^2/D(\sigma)$ and, upon the rescaling $T \to T/\lambda^2$, Eq.~\eqref{eq:final} yields for the connected part of the \threetothree amplitude
\begin{align}
\langle q_1,q_2,q_3|\hat T_c(s)| p_1,p_2,p_3\rangle  =
&\frac{1}{3!}\sum_{n=1}^3\sum_{m=1}^3 T_{22}(\sigma(q_n))
\langle q_n|T(s)|p_m\rangle
T_{22}(\sigma(p_m)) 
\end{align}
with 
\newpage
\begin{align}
\label{eq:20}
\langle q| T(s)|p\rangle  = \langle q| C(s)|p\rangle +&
\frac{1}{m^2-(P-p-q)^2-i\epsilon}\\\nonumber
&\qquad-
\int
\frac{\mathrm{d}^3\boldsymbol{\ell}}{(2\pi)^3}
\frac{1}{2E_\ell}
T_{22}(\sigma(\ell))\left(\langle \ell| C(s)|p\rangle +  \frac{1}{m^2-(P-p-\ell)^2-i\epsilon}\right)\langle \ell| T(s)|p\rangle\,,
\end{align}
where, again, $E_\ell^2=(\boldsymbol{\ell}^2+m^2)$, all involved particles are on-energy-shell, and $\langle q_n|C(s)|p_m\rangle$ represents an unknown function with no imaginary part in the physical region. For example, $C(s)$ can generate the real term in Eq.~\eqref{eq:B}. Notably, all dependence on the details on the parametrization of $T_{22}$ has disappeared; also, in this form it becomes clear that only the on-shell \twototwo amplitude enters the integral equation. For physical \threetothree scattering, the $\boldsymbol{\ell}$ integration scans $T_{22}$ from a maximal $\sigma(\ell)=(\sqrt{s}-m)^2$ down to threshold at $\sigma(\ell)=(2m)^2$. For larger values of $|\boldsymbol{\ell}|$, $T_{22}$ is still evaluated on-shell but below threshold. The \threetothree scattering equation requires, thus, not only the physical on-shell \twototwo scattering amplitude (or, equivalently, the phase shift), but also its extrapolation below threshold.

\begin{figure}
\begin{center}
\includegraphics[width=\textwidth]{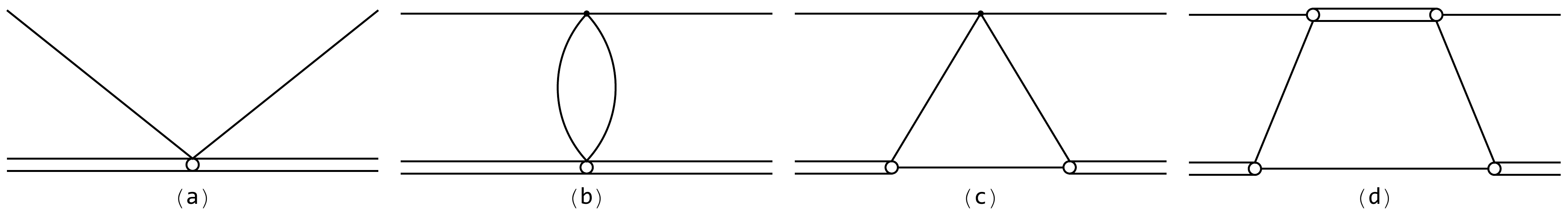}
\end{center}
\caption{Additional contributions to the isobar-spectator interaction. Figure (a) shows a contribution from three-body forces, while (b) shows a loop contributions without imaginary part in the $s$-channel physical region. The diagram in (c) has a nonzero imaginary part, but it contains a non-consistent treatment of the \twototwo scattering amplitude (the pointlike interaction in the top vertex). One has to replace this with diagram (d), which is already generated by the BSE. See text for further discussion.}
\label{fig:B_extra}
\end{figure}

Finally, one may ask what other contributions can be added to $B$, or similarly what enters $C(s)$ in Eq.~\eqref{eq:20}, without violating unitarity. For example, let us consider the diagrams in Fig.~\ref{fig:B_extra}. The diagram in (a) contains a pointlike real isobar-spectator interaction. We can think of this term as the exchange of an infinitely heavy particle in the crossed channel, which generates a left hand cut at infinity, but does not spoil unitarity. An alternative interpretation would be that of a genuine three-body force that cannot be accommodated by exchange of the light particle; through iteration in the scattering equation, such a contribution could lead to the formation of a three-body resonance.
Similarly, the diagram in (b) contains only $t$-channel singularities which appear as left hand cuts in the $s$-channel, and do not contribute to the unitarity equation. Conversely, the diagram in (c) actually has an imaginary part in the physical region: The three mesons in the intermediate state can go simultaneously on-shell, giving a singular contribution in $s$. Therefore, this diagram cannot be added to $B$. The top vertex describes a point-like spectator-spectator interaction, but the only consistent treatment of the \twototwo scattering within the isobar approximation is via the exchange of an isobar. A consistent treatment of the top vertex should therefore look like the diagram in (d), which is already generated via the BSE. 

In summary, if one aims at a microscopic description in terms of Feynman diagrams, the kernel of the BSE can be expanded with contributions that cannot go on-shell in the physical region; here, we prefer to think in terms of unitary sub-amplitudes and identify the imaginary parts necessary to maintain three-body unitarity; the necessary degrees of freedom to analyze data in future applications comes rather from real isobar-spectator contact terms and the freedom in the choice of the isobar propagator $S(\sigma(k))$.

\section{Summary and Conclusions}\label{sec:conclusions}

Addressing three-body unitarity for the \threetothree scattering amplitude in re-scattering schemes, such as provided by the Bethe-Salpeter equation (BSE),  is considerably more intricate than in the two-body case. The Amado model provides an appealing solution to the problem because it can be reduced to an integral equation in one three-momentum by using the isobar assumption. However, a closer inspection of three-body unitarity reveals that there are two shortcomings in the original proof by Aaron, Amado, and Young: Unitarity is only formulated in terms of asymptotically stable states, not isobars, and, second, certain imaginary parts of the Bethe-Salpeter equation do not vanish  above break-up. Due to these problems, the validity of the original proof is restricted to scattering energies below break-up and to bound-state particle scattering.

In this study, we have consequently formulated the scattering problem  in terms of asymptotically stable states, including disconnected diagrams. As a result, the right-hand side of the unitarity relation acquires several new terms. These terms correspond exactly to the problematic extra terms for the imaginary parts of the Bethe-Salpeter equation. The obtained matching relations to derive the driving term and the Green's function of the BSE are consistent and lead to a scattering equation very similar to the original results by Aaron, Amado, and Young. However, the scattering equation derived here is manifestly three-body unitary also above breakup. In the absence of three-body forces one can recast it in terms of on-shell \twototwo scattering amplitudes. In future work, the formalism will be generalized to the case of several isobars, the inclusion of 2-body channels, and (iso)spin for particles and isobars.

\begin{acknowledgments}

This work is supported by the National Science Foundation (CAREER grant
PHY-1452055, NSF/PIF grant No. 1415459), by GWU (startup grant), and by the U.S. Department of Energy, Office of Science, Office of Nuclear Physics under contract DE-AC05-06OR23177.
M.D. is also supported by the U.S. Department of Energy, Office of Science,
Office of Nuclear Physics under grant No. DE-SC001658. M.M. is thankful to the German Research Foundation (DFG) for the financial support, under the fellowship MA 7156/1-1, as well as George Washington University for hospitality and inspiring environment.
\end{acknowledgments}

\appendix
\section{Analytic expressions for the matching process}
\label{sec:appendix_a}

First we compare the topologies, which do not contain driving term of the BSE~\eqref{eq:bse1}, \ie (1a), (1b) and (4a). For diagram (1a), \ie, the right-hand side of the unitarity relation, we obtain:
\begin{align}\label{app:unit1a}
\langle q_1,q_2,q_3|&(\hat T-\hat T^{\dagger})| p_1,p_2,p_3\rangle_{{\rm (1a)}}\\\nonumber
=&i\int\prod_{\ell=1}^3\left[\frac{\mathrm{d}^4k_\ell}{(2\pi)^{4}}\,(2\pi)\delta^+(k_\ell^2-m^2)\right]\,
(2\pi)^4\delta^4\left(P-\sum_{\ell=1}^3\,k_\ell\right)
\\\nonumber
&
\times\frac{1}{3!}\sum_{n,i}v(q_{\bar{n}},q_{\barr{n}})S^\dagger (\sigma(q_n))v(k_{\bar{i}},k_{\barr{i}})\,
2E_{\mathbf{q}_n}(2\pi)^3\delta^3(\mathbf{q}_n-\mathbf{k}_i) 
\\&
\times
\frac{1}{3!}\sum_{j,m} v(k_{\bar{j}},k_{\barr{j}})S(\sigma(k_j))\langle k_j|T|p_m\rangle S(\sigma(p_m))v(p_{\bar{m}},p_{\barr{m}}) \times \delta_{ij}\nonumber\\\nonumber
=&i\frac{1}{3!}\,\sum_{n,m,i}\frac{1}{3!}
\int\prod_{\ell=1}^3\left[\frac{\mathrm{d}^4k_\ell}{(2\pi)^{4}}\,(2\pi)\delta^+(k_\ell^2-m^2)\right]\,
(2\pi)^4\delta^4\left(P-\sum_{\ell=1}^3\,k_\ell\right)\\\nonumber
&\times v(q_{\bar{n}},q_{\barr{n}})S^\dagger (\sigma(q_n))v(k_{\bar{i}},k_{\barr{i}})
\,
2E_{\mathbf{q}_n}(2\pi)^3\delta^3(\mathbf{q}_n-\mathbf{k}_i) 
\times
 v(k_{\bar{i}},k_{\barr{i}})S(\sigma(k_i))\langle k_i|T|p_m\rangle S(\sigma(p_m))v(p_{\bar{m}},p_{\barr{m}})\\\nonumber
=&i\frac{1}{3!}\sum_{n,m,i}\frac{1}{3!}
\int\prod_{\ell\ne i}\left[\frac{\mathrm{d}^4k_\ell}{(2\pi)^{4}}\,(2\pi)\delta^+(k_\ell^2-m^2)\right]\,
(2\pi)^4\delta^4\left(P-\sum_{\ell\ne i}\,k_\ell-q_n\right)
\\\nonumber&
\times
v(q_{\bar{n}},q_{\barr{n}})S^\dagger (\sigma(q_n))v(k_{\bar{i}},k_{\barr{i}})\times
v(k_{\bar{i}},k_{\barr{i}})S(\sigma(q_n))\langle q_n|T|p_m\rangle S(\sigma(p_m))v(p_{\bar{m}},p_{\barr{m}})\\\nonumber
=&i\frac{1}{3!}\sum_{n,m}v(q_{\bar{n}},q_{\barr{n}})S^\dagger (\sigma(q_n))
S(\sigma(q_n))\langle q_n|T|p_m\rangle S(\sigma(p_m))v(p_{\bar{m}},p_{\barr{m}})\\\nonumber
&\times
\frac{1}{2}
\int\prod_{\ell\ne 1}\left[\frac{\mathrm{d}^4k_\ell}{(2\pi)^{4}}\,(2\pi)\delta^+(k_\ell^2-m^2)\right]\,
(2\pi)^4\delta^4\left(P-\sum_{\ell\ne 1}\,k_\ell-q_n\right)
\times
\left(v(k_{2},k_{3})\right)^2\\\nonumber
=&i\frac{1}{3!}\sum_{n,m}v(q_{\bar{n}},q_{\barr{n}})
S^\dagger (\sigma(q_n))
S(\sigma(q_n))
\langle q_n|T|p_m\rangle 
S(\sigma(p_m))
v(p_{\bar{m}},p_{\barr{m}})\\\nonumber
&\times
\frac{1}{2}
\int\frac{\mathrm{d}^4\bar K\mathrm{d}^4\bar P}{(2\pi)^{8}}\,(2\pi)^2\delta^+\left(\left(\frac{\bar P+2\bar K}{2}  \right)^2-m^2\right)\delta^+\left(\left(\frac{\bar P-2\bar K}{2}  \right)^2-m^2\right)\\\nonumber
&~~~~~~~~~~~~~~~~~~~~~\times(2\pi)^4\delta^4\left(P-\bar P-q_n\right)
\left(v\left(\frac{\bar P+2\bar K}{2},\frac{\bar P-2\bar K}{2}\right)\right)^2\\\nonumber
=&
i\frac{1}{3!}\sum_{n,m}v(q_{\bar{n}},q_{\barr{n}})S^\dagger (\sigma(q_n))
S(\sigma(q_n))\langle q_n|T|p_m\rangle S(\sigma(p_m))v(p_{\bar{m}},p_{\barr{m}})\\\nonumber
&\times
\frac{1}{2(2\pi)^2}
\int \mathrm{d}^4\bar K\,\delta^+\left(\left(\frac{P-q_n+2\bar K}{2}  \right)^2-m^2\right)\delta^+\left(\left(\frac{P-q_n-2\bar K}{2}  \right)^2-m^2\right)
\left(v\left(\frac{P-q_n+2\bar K}{2},\frac{P-q_n-2\bar K}{2}\right)\right)^2\,,
\end{align}
where the non-trivial steps are: 1) In first equality, the factor $\delta_{ij}$ indicates that the spectator is the same for both $T_{c/d}$ matrices \ie, one obtains a self-energy-type of geometry, and {\it not} an exchange-type combination of intermediate particles; 2) In the third equality we have used $(2\pi)\delta^+(k_1^2-m^2)2E_{\mathbf{q}_n}(2\pi)^3\delta^3(\mathbf{q}_n-\mathbf{k}_1)=(2\pi)^4\delta^4(k_1-q_n)$;
3) In the fifth equality we have transformed the integration via $2\bar K:=k_2-k_3$ and $\bar P:=k_2+k_3$.

On the other hand, for the discontinuity of the same part of the scattering matrix, evaluated using the BSE Ansatz, we obtain
\begin{align}\label{app:bse1a}
\langle q_1,q_2,q_3|&(\hat T-\hat T^{\dagger})| p_1,p_2,p_3\rangle_{{\rm (1a)}}
=
\frac{1}{3!}\sum_{n,m}
v(q_{\bar{n}},q_{\barr{n}})
(S(\sigma(q_n))-S^\dagger (\sigma(q_n)))
\langle q_n|T|p_m\rangle 
S(\sigma(p_m))v(p_{\bar{m}},p_{\barr{m}})\,,
\end{align}

Comparing now the latter two equations for any given in- and outgoing external momenta $(p_1,p_2,p_3)$ and $(q_1,q_2,q_3)$, respectively, we arrive at the following matching/discontinuity relation for the propagator of the isobar:
\begin{align}
\label{eq:result1a}
S(\sigma(q_n))-S^\dagger &(\sigma(q_n))=i\frac{S(\sigma(q_n))S^\dagger (\sigma(q_n))}{2(2\pi)^2}\\\nonumber
&\times\int \mathrm{d}^4\bar K\,\delta^+\left(\left(\frac{P-q_n+2\bar K}{2}  \right)^2-m^2\right)\delta^+\left(\left(\frac{P-q_n-2\bar K}{2}  \right)^2-m^2\right)
\left(v\left(\frac{P-q_n+2\bar K}{2},\frac{P-q_n-2\bar K}{2})\right)\right)^2\,.
\end{align}

In the same manner we obtain from the comparison of diagrams of type (1b)
\begin{align}
\label{eq:result1b}
S(\sigma(p_m))-S^\dagger &(\sigma(p_m))=i\frac{S(\sigma(p_m))S^\dagger (\sigma(p_m))}{2(2\pi)^2}\\\nonumber
\times&\int \mathrm{d}^4\bar K\,\delta^+\left(\left(\frac{P-p_m+2\bar K}{2}  \right)^2-m^2\right)\delta^+\left(\left(\frac{P-p_m-2\bar K}{2}  \right)^2-m^2\right)
\left(v\left(\frac{P-p_m+2\bar K}{2},\frac{P-p_m-2\bar K}{2}\right)\right)^2\,.
\end{align}

The diagram of type (4a) on the left panel of Fig.~\ref{fig:compare} reads
\begin{align}
\label{eq:unit4a}
\langle q_1,q_2,q_3|&(\hat T-\hat T^{\dagger})| p_1,p_2,p_3\rangle_{{\rm (4a)}}\\\nonumber	
=&i\int\prod_{\ell=1}^3\left[\frac{\mathrm{d}^4k_\ell}{(2\pi)^{4}}(2\pi)\delta^+(k_\ell^2-m^2)\right]\,
(2\pi)^4\delta^4\left(P-\sum_{\ell=1}^3\,k_\ell\right)\\\nonumber
&\times\frac{1}{3!}\sum_{n,i}v(q_{\bar{n}},q_{\barr{n}})S^\dagger (\sigma(q_n))\langle q_n|T^{\dagger}|k_i\rangle S^\dagger (\sigma(k_i))v(k_{\bar{i}},k_{\barr{i}})\nonumber \\
&\times
\frac{1}{3!}\sum_{j,m} v(k_{\bar{j}},k_{\barr{j}})S(\sigma(k_j))\langle k_j|T|p_m\rangle S(\sigma(p_m))v(p_{\bar{m}},p_{\barr{m}})\times \delta_{ij}\nonumber\\
\nonumber
=&i\int\prod_{\ell=1}^3\left[\frac{\mathrm{d}^4k_\ell}{(2\pi)^{4}}(2\pi)\delta^+(k_\ell^2-m^2)\right]\,
(2\pi)^4\delta^4\left(P-\sum_{\ell=1}^3\,k_\ell\right)\\\nonumber
&\times\frac{1}{3!}\sum_{n,m}v(q_{\bar{n}},q_{\barr{n}})S^\dagger (\sigma(q_n))S(\sigma(p_m))v(p_{\bar{m}},p_{\barr{m}}) 
\times\frac{1}{3!}\sum_{i}\langle q_n|T^{\dagger}|k_i\rangle S^\dagger (\sigma(k_i))v(k_{\bar{i}},k_{\barr{i}})^2 S(\sigma(k_i))\langle k_i|T|p_m\rangle  \\\nonumber
=&i\frac{1}{3!}\sum_{n,m}v(q_{\bar{n}},q_{\barr{n}})S^\dagger (\sigma(q_n))S(\sigma(p_m))v(p_{\bar{m}},p_{\barr{m}}) \\\nonumber
&\times\int\prod_{\ell\neq 1}\left[\frac{\mathrm{d}^4k_\ell}{(2\pi)^{4}}(2\pi)\delta^+(k_\ell^2-m^2)\right]\,
(2\pi)^4\delta^4\left(P-\sum_{\ell\neq 1}k_\ell-k_1\right) v(k_2,k_3)^2 
\\\nonumber&
\times\frac{1}{2}\int\frac{\mathrm{d}^4k_1}{(2\pi)^{4}}\langle q_n|T^{\dagger}|k_1\rangle S^\dagger (\sigma(k_1))S(\sigma(k_1))\langle k_1|T|p_m\rangle  (2\pi)\delta^+(k_1^2-m^2)
\\\nonumber
=&i\frac{1}{3!}\sum_{n,m}v(q_{\bar{n}},q_{\barr{n}})
S^\dagger (\sigma(q_n))
S(\sigma(p_m))v(p_{\bar{m}},p_{\barr{m}}) \\\nonumber
&\times\int\frac{\mathrm{d}^4\bar K \mathrm{d}^4\bar P}{(2\pi)^{8}}(2\pi)^2\delta^+\left(\left(\frac{\bar P+2\bar K}{2}\right)^2-m^2\right)\delta^+\left(\left(\frac{\bar P-2\bar K}{2}  \right)^2-m^2\right) 
\left(v\left(\frac{\bar P+2\bar K}{2},\frac{\bar P-2\bar K}{2}\right)\right)^2
\\\nonumber
&\times\frac{1}{2}\int\frac{\mathrm{d}^4k_1}{(2\pi)^{4}}\langle q_n|T^{\dagger}|k_1\rangle S^\dagger (\sigma(k_1)) S(\sigma(k_1))\langle k_1|T|p_m\rangle  
(2\pi)\delta^+(k_1^2-m^2)(2\pi)^4\delta^4\left(P-\bar P-k_1\right)\\\nonumber
=&i\frac{1}{3!}\sum_{n,m}v(q_{\bar{n}},q_{\barr{n}})S^\dagger (\sigma(q_n))S(\sigma(p_m))v(p_{\bar{m}},p_{\barr{m}}) 
\times\int\frac{\mathrm{d}^4k_1}{(2\pi)^{4}}(2\pi)\delta^+(k_1^2-m^2)\langle q_n|T^{\dagger}|k_1\rangle S^\dagger (\sigma(k_1))S(\sigma(k_1))\langle k_1|T|p_m\rangle \\\nonumber
&\times\frac{1}{2}\int\frac{\mathrm{d}^4\bar K \mathrm{d}^4\bar P}{(2\pi)^{8}}
\left(v\left(\frac{\bar P+2\bar K}{2},\frac{\bar P-2\bar K}{2}\right)\right)^2
(2\pi)^2\delta^+\left(\left(\frac{\bar P+2\bar K}{2}\right)^2-m^2\right)\delta^+\left(\left(\frac{\bar P-2\bar K}{2}  \right)^2-m^2\right) 
\\\nonumber&
\times(2\pi)^4\delta^4\left(P-\bar P-k_1\right) \\\nonumber
=&i\frac{1}{3!}\sum_{n,m}v(q_{\bar{n}},q_{\barr{n}})S^\dagger (\sigma(q_n))S(\sigma(p_m))v(p_{\bar{m}},p_{\barr{m}}) 
\times\int\frac{\mathrm{d}^4k_1}{(2\pi)^{4}}(2\pi)\delta^+(k_1^2-m^2)\langle q_n|T^{\dagger}|k_1\rangle S^\dagger (\sigma(k_1))S(\sigma(k_1))\langle k_1|T|p_m\rangle \\\nonumber
&\times\frac{1}{2}\int\frac{\mathrm{d}^4\bar K}{(2\pi)^{4}}
\left(v\left(\frac{P - k_1+2\bar K}{2},\frac{P - k_1-2\bar K}{2}\right)\right)^2
(2\pi)^2\delta^+\left(\left(\frac{P - k_1 +2\bar K}{2}\right)^2-m^2\right)\delta^+\left(\left(\frac{P - k_1 -2\bar K}{2}  \right)^2-m^2\right)\,,
\end{align}
while the discontinuity of the corresponding topology on the right panel of Fig.~\ref{fig:compare} reads
\begin{align}
\label{eq:bse4a}
\langle q_1,q_2,q_3|&(\hat T-\hat T^{\dagger})| p_1,p_2,p_3\rangle_{{\rm (4a)}}\\\nonumber	
=&\frac{1}{3!}\sum_{n,m}\int\frac{\mathrm{d}^4 k}{(2\pi)^4}v(q_{\bar{n}},q_{\barr{n}})S^\dagger (\sigma(q_n))\langle q_n|T^{\dagger}|k\rangle
(\tau (\sigma(k))-\tau^{\dagger}(\sigma(k))) 
\langle k|T|p_m\rangle S(\sigma(p_m))v(p_{\bar{m}},p_{\barr{m}})\\\nonumber
=&\frac{1}{3!}\sum_{n,m}v(q_{\bar{n}},q_{\barr{n}})S^\dagger (\sigma(q_n))S(\sigma(p_m))v(p_{\bar{m}},p_{\barr{m}}) 
\times\int\frac{\mathrm{d}^4 k_1}{(2\pi)^4}\langle q_n|T^{\dagger}|k_1\rangle
(\tau (\sigma(k_1))-\tau^{\dagger}(\sigma(k_1))) \langle k_1|T|p_m\rangle\,. \hspace{15cm}
\end{align}

Since the last two equations hold for any choice of external in- and outgoing  momenta, the element-wise comparison yields
\begin{align}
\label{eq:result4a}
S(\sigma(k))-&S^\dagger (\sigma(k)) = \frac{i}{2(2\pi)^{2}}S^\dagger (\sigma(k))S(\sigma(k)) \\\nonumber
&\times\int\mathrm{d}^4\bar K
\left(v\left(\frac{P - k+2\bar K}{2},\frac{P - k-2\bar K}{2}\right)\right)^2
\delta^+\left(\left(\frac{P - k +2\bar K}{2}\right)^2-m^2\right)\delta^+\left(\left(\frac{P - k -2\bar K}{2}  \right)^2-m^2\right)\,,
\end{align}
when using the condition ${\tau(\sigma(k)) = (2\pi)\delta^+(k^2-m^2)S(\sigma(k))}$, chosen to impose consistency of all three relations (\eqref{eq:result1a}, \eqref{eq:result1b} and \eqref{eq:result4a}) for the isobar propagator discontinuity.

The remaining topologies are (2), (3a), (3b), (4b), which are obviously related to the discontinuity of the driving term of the BSE \eqref{eq:bse3}. For the diagram of type (2) on the left panel of Fig.~\ref{fig:compare} we obtain
\begin{align}
\label{eq:unit2}
\langle q_1,q_2,q_3|&(\hat T-\hat T^{\dagger})| p_1,p_2,p_3\rangle_{{\rm (2)}}\\\nonumber
=&i\int\prod_{\ell=1}^3\left[\frac{\mathrm{d}^4k_\ell}{(2\pi)^{4}}(2\pi)\delta^+(k_\ell^2-m^2)\right]\,
(2\pi)^4\delta^4\left(P-\sum_{\ell=1}^3k_\ell\right)\\\nonumber
&\times\frac{1}{3!}\sum_{n,i}
v(q_{\bar{n}},q_{\barr{n}})S^\dagger (\sigma(q_n))
v(k_{\bar{i}},k_{\barr{i}})2E_{\mathbf{q}_n}
(2\pi)^3\delta^3(\mathbf{q}_n-\mathbf{k}_i) \nonumber \\
&\times
\frac{1}{3!}\,\sum_{j,m} 
v(k_{\bar{j}},k_{\barr{j}})
S(\sigma(p_m))
v(p_{\bar{m}},p_{\barr{m}})2E_{\mathbf{p}_m}(2\pi)^3\delta^3(\mathbf{p}_m-\mathbf{k}_j)\times(1-\delta_{ij})\nonumber\\\nonumber
=&i\frac{1}{3!}\frac{1}{3!}\sum_{n,m}
v(q_{\bar{n}},q_{\barr{n}})S^\dagger (\sigma(q_n))S(\sigma(p_m))
v(p_{\bar{m}},p_{\barr{m}})\\\nonumber
&\times 3\int\frac{\mathrm{d}^4k_{1}}{(2\pi)^{4}}\,(2\pi)\delta^+(k_1^2-m^2)\\\nonumber
&\times 2\int\frac{\mathrm{d}^4k_{2}}{(2\pi)^{4}}
(2\pi)\delta^+(k_2^2-m^2)
v(k_2,k_1)
2E_{\mathbf{q}_n}(2\pi)^3\delta^3(\mathbf{q}_n-\mathbf{k}_2) 
\\\nonumber
&\times\int\frac{\mathrm{d}^4k_{3}}{(2\pi)^{4}}(2\pi)\delta^+(k_3^2-m^2)v(k_3,k_1)2E_{\mathbf{p}_m}(2\pi)^3\delta^3(\mathbf{p}_m-\mathbf{k}_3) 
\times(2\pi)^4\delta^4\left(P-k_1-k_2-k_3\right)\\\nonumber
=&i\frac{1}{3!}\sum_{n,m}
v(q_{\bar{n}},q_{\barr{n}})
S^\dagger (\sigma(q_n))S(\sigma(p_m))
v(p_{\bar{m}},p_{\barr{m}})\int\frac{\mathrm{d}^4k_1}{(2\pi)^{4}}(2\pi)\delta^+(k_1^2-m^2) \\\nonumber
&\times\int\frac{\mathrm{d}^4k_2}{(2\pi)^{4}}
v(k_2,k_1)
(2\pi)^4\delta^4(q_n-k_2) \\\nonumber
&\times\int\frac{\mathrm{d}^4k_3}{(2\pi)^{4}}
v(k_3,k_1)
(2\pi)^4\delta^4(p_m-k_3)
(2\pi)^4\delta^4\left(P-k_1-k_2-k_3\right) \\\nonumber
=&i\frac{1}{3!}\sum_{n,m}
v(q_{\bar{n}},q_{\barr{n}})
S^\dagger (\sigma(q_n))
S(\sigma(p_m))
v(p_{\bar{m}},p_{\barr{m}}) 
\times\int\frac{\mathrm{d}^4k_1}{(2\pi)^{4}}
(2\pi)\delta^+(k_1^2-m^2)
(2\pi)^4\delta^4\left(P-k_1-q_n-p_m\right)
v(q_n,k_1)
v(p_m,k_1) \\\nonumber
=&i\frac{1}{3!}\sum_{n,m}
v(q_{\bar{n}},q_{\barr{n}})
S^\dagger (\sigma(q_n))
S(\sigma(p_m))
v(p_{\bar{m}},p_{\barr{m}})
\times 
v(q_n,P-q_n-p_m)
(2\pi)\delta^+((P-q_n-p_m)^2-m^2)
v(p_m, P-q_n-p_m)\,.
\end{align}
Whereas, imposing BSE Ansatz the same discontinuity yields 
\begin{align}
\label{eq:bse22}
\langle q_1,q_2,q_3|(\hat T-\hat T^{\dagger})| p_1,p_2,p_3\rangle_{{\rm (2)}}
=\frac{1}{3!}\sum_{n,m}
v(q_{\bar{n}},q_{\barr{n}})
S^\dagger (\sigma(q_n))
(\langle q_n|B|p_m \rangle - \langle q_n|B^\dagger|p_m \rangle)
S(\sigma(p_m))
v(p_{\bar{m}},p_{\barr{m}})
\,.
\end{align}
We can set the latter both equations equal and compare individual terms in the sums over $m$ and $n$, since they both hold for any external momenta. Therefore, the discontinuity of the driving term of the BSE equations for on-shell momenta reads
\begin{align}
\label{eq:result2}
\langle q_n|B|p_m \rangle - \langle q_n|B^\dagger|p_m \rangle 
 = iv((P-2q_n-p_m)^2)(2\pi)\delta^+((P-q_n-p_m)^2-m^2)v((P-q_n-2p_m)^2)\,.
\end{align}

The diagram of type (3a) on the left panel of Fig.~\ref{fig:compare} reads
\begin{align}
\label{eq:unit3a}
\langle q_1,q_2,q_3|&(\hat T-\hat T^{\dagger})| p_1,p_2,p_3\rangle_{{\rm (3a)}}\\\nonumber
=&i\int\prod_{\ell=1}^3\left[\frac{\mathrm{d}^4k_\ell}{(2\pi)^{4}}\,(2\pi)\delta^+(k_\ell^2-m^2)\right]\,
(2\pi)^4\delta^4\left(P-\sum_{\ell=1}^3k_\ell\right)
\\\nonumber
&\times\frac{1}{3!}\sum_{n,i}
v(q_{\bar{n}},q_{\barr{n}})
S^\dagger (\sigma(q_n))
v(k_{\bar{i}},k_{\barr{i}})
2E_{\mathbf{q}_n}(2\pi)^3\delta^3(\mathbf{q}_n-\mathbf{k}_i) 
\times
\frac{1}{3!}\,\sum_{j,m} 
v(k_{\bar{j}},k_{\barr{j}})
S(\sigma(k_3))
\langle k_3|T|p_m\rangle 
S(\sigma(p_m))
v(p_{\bar{m}},p_{\barr{m}})\\\nonumber
&\times (1-\delta_{ij}) \\\nonumber
=&i\frac{1}{3!}\frac{1}{3!}\sum_{n,m}
v(q_{\bar{n}},q_{\barr{n}})
S^\dagger (\sigma(q_n))S(\sigma(p_m))
v(p_{\bar{m}},p_{\barr{m}})\\\nonumber
&\times3\int\frac{\mathrm{d}^4k_1}{(2\pi)^{4}}\,
(2\pi)\delta^+(k_1^2-m^2)\\\nonumber
&\times2\int\frac{\mathrm{d}^4k_2}{(2\pi)^{4}}
(2\pi)\delta^+(k_2^2-m^2)
v(k_2,k_1)
2E_{\mathbf{q}_n}
(2\pi)^3\delta^3(\mathbf{q}_n-\mathbf{k}_i) \\\nonumber
&\times\int\frac{\mathrm{d}^4k_3}{(2\pi)^{4}}
(2\pi)\delta^+(k_3^2-m^2)
v(k_3,k_1)
S(\sigma(k_3))
\langle k_3|T|p_m\rangle \\\nonumber
&\times(2\pi)^4\delta^4\left(P-k_1-k_2-k_3\right) \\\nonumber
=&i\frac{1}{3!}\sum_{n,m}
v(q_{\bar{n}},q_{\barr{n}})
S^\dagger (\sigma(q_n))
S(\sigma(p_m))
v(p_{\bar{m}},p_{\barr{m}})\\\nonumber
&
\times\int\frac{\mathrm{d}^4k_1}{(2\pi)^{4}}
(2\pi)\delta^+(k_1^2-m^2)\int\frac{\mathrm{d}^4k_2}{(2\pi)^{4}}
v(k_2,k_1)
(2\pi)^4\delta^4(q_n-k_2) \\\nonumber
&\times\int\frac{\mathrm{d}^4k_3}{(2\pi)^{4}}
(2\pi)\delta^+(k_3^2-m^2)
v(k_3,k_1)
S(\sigma(k_3))\langle k_3|T|p_m\rangle
(2\pi)^4\delta^4\left(P-k_1-k_2-k_3\right) \\\nonumber
=&i\frac{1}{3!}\sum_{n,m}
v(q_{\bar{n}},q_{\barr{n}})
S^\dagger (\sigma(q_n))
S(\sigma(p_m))
v(p_{\bar{m}},p_{\barr{m}})\\\nonumber
&\times\int\frac{\mathrm{d}^4k_1}{(2\pi)^{4}}
(2\pi)\delta^+(k_1^2-m^2)
v(q_n,k_1) 
\times 
\int\frac{\mathrm{d}^4k_3}{(2\pi)^{4}}
(2\pi)\delta^+(k_3^2-m^2)
v(k_3,k_1)
S(\sigma(k_3))
\langle k_3|T|p_m\rangle
(2\pi)^4\delta^4\left(P-k_1-q_n-k_3\right) \\\nonumber
=&i\frac{1}{3!}\sum_{n,m}
v(q_{\bar{n}},q_{\barr{n}})S^\dagger (\sigma(q_n))
S(\sigma(p_m))
v(p_{\bar{m}},p_{\barr{m}}) \\\nonumber
&\times \int\frac{\mathrm{d}^4k_3}{(2\pi)^{4}}
(2\pi)\delta^+(k_3^2-m^2)
S(\sigma(k_3))
\langle k_3|T|p_m\rangle 
\times \int\frac{\mathrm{d}^4k_1}{(2\pi)^{4}}
v(k_3,k_1)
(2\pi)\delta^+(k_1^2-m^2)
v(q_n,k_1)
(2\pi)^4\delta^4\left(P-k_1-q_n-k_3\right) \\\nonumber
=&i\frac{1}{3!}\sum_{n,m}v(q_{\bar{n}},q_{\barr{n}})S^\dagger (\sigma(q_n))S(\sigma(p_m))v(p_{\bar{m}},p_{\barr{m}}) \\\nonumber
&\times \int\frac{\mathrm{d}^4k_3}{(2\pi)^{4}}(2\pi)\delta^+(k_3^2-m^2)S(\sigma(k_3))\langle k_3|T|p_m\rangle
\times v(P-q_n-k_3,k_3)(2\pi)\delta^+((P-q_n-k_3)^2-m^2)v(q_n,P-q_n-k_3)\,.
\end{align}
The corresponding topology on the right panel of Fig.~\ref{fig:compare} reads 
\begin{align}
\label{eq:bse33}
\langle q_1,q_2,q_3|&(\hat T-\hat T^{\dagger})| p_1,p_2,p_3\rangle_{{\rm (3a)}}	\\\nonumber
=&\frac{1}{3!}\sum_{n,m}
v(q_{\bar{n}},q_{\barr{n}})
S^\dagger (\sigma(q_n))
\int\frac{\mathrm{d}^4k_3}{(2\pi)^{4}}
(\langle q_n|B|k_3 \rangle - \langle q_n|B^\dagger|k_3 \rangle)
\tau(\sigma_{k_3})
\langle k_3|T|p_m \rangle 
S(\sigma(p_m))
v(p_{\bar{m}},p_{\barr{m}}) 
\\\nonumber
=&\frac{1}{3!}\sum_{n,m}
v(q_{\bar{n}},q_{\barr{n}})
S^\dagger (\sigma(q_n))
S(\sigma(p_m))
v(p_{\bar{m}},p_{\barr{m}}) 
\times\int\frac{\mathrm{d}^4k_3}{(2\pi)^{4}}
\tau (\sigma_{k_3})
\langle k_j|T|p_m \rangle
(\langle q_n|B|k_3 \rangle - \langle q_n|B^\dagger|k_3 \rangle) \ .
\hspace{15cm}
\end{align}
Comparison of the last two equations leads to the same type of matching relation for the driving term of the BSE as before when imposing ${\tau(\sigma(k)) = (2\pi)\delta^+(k^2-m^2)S(\sigma(k))}$. However, this time the equation holds for any values of $k_3$ as long as $k^2=m^2$
\begin{align}
\label{eq:result3a}
\langle q_n|B|k_3 \rangle - \langle q_n|&B^\dagger|k_3 \rangle= 
   iv(P-q_n-k_3,k_3)(2\pi)\delta^+((P-q_n-k_3)^2-m^2)v(q_n,P-q_n-k_3)\,.
\end{align}
The same kind of matching relation, but for the incoming momentum being on the mass shell  can be derived along the same lines from the comparison of the topologies (3b) in Fig.~\ref{fig:compare}.

Finally, the diagram of type (4b) on the left panel of Fig.~\ref{fig:compare} reads
\begin{align}
\label{eq:unit4b}
\langle q_1,q_2,q_3|&(\hat T-\hat T^{\dagger})| p_1,p_2,p_3\rangle_{{\rm (4b)}}\\\nonumber	
=&i\int\prod_{\ell=1}^3\left[\frac{\mathrm{d}^4k_\ell}{(2\pi)^{4}}\,(2\pi)\delta^+(k_\ell^2-m^2)\right]\,
(2\pi)^4\delta^4\left(P-\sum_{\ell=1}^3k_\ell\right)\\\nonumber
&\times\frac{1}{3!}\sum_{n,i}v(q_{\bar{n}},q_{\barr{n}})S^\dagger (\sigma(q_n))\langle q_n|T^{\dagger}|k_i\rangle S^\dagger (\sigma(k_i))v(k_{\bar{i}},k_{\barr{i}}) 
\times
\frac{1}{3!}\,\sum_{j,m} v(k_{\bar{j}},k_{\barr{j}})S(\sigma(k_j))\langle k_j|T|p_m\rangle S(\sigma(p_m))v(p_{\bar{m}},p_{\barr{m}})\\\nonumber
&\times (1-\delta_{ij}) \\\nonumber
=&i\frac{1}{3!}\frac{1}{3!}\sum_{n,m}
v(q_{\bar{n}},q_{\barr{n}})
S^\dagger (\sigma(q_n))
S(\sigma(p_m))v(p_{\bar{m}},p_{\barr{m}}) \\\nonumber
&\times3\int\frac{\mathrm{d}^4k_1}{(2\pi)^{4}}
(2\pi)\delta^+(k_1^2-m^2)
S^\dagger (\sigma(k_1))
\langle q_n|T^{\dagger}|k_1\rangle \\\nonumber
&\times2\int\frac{\mathrm{d}^4k_2}{(2\pi)^{4}}
(2\pi)\delta^+(k_2^2-m^2)
S(\sigma(k_2))
\langle k_2|T|p_m\rangle \\\nonumber
&\times\int\frac{\mathrm{d}^4k_3}{(2\pi)^{4}}
v(k_1,k_3)(2\pi)\delta^+(k_3^2-m^2)
v(k_2,k_3)
(2\pi)^4\delta^4\left(P-k_3-k_1-k_2\right) \\\nonumber
=&i\frac{1}{3!}\sum_{n,m}v(q_{\bar{n}},q_{\barr{n}})S^\dagger (\sigma(q_n))S(\sigma(p_m))v(p_{\bar{m}},p_{\barr{m}}) \\\nonumber
&\times\int\frac{\mathrm{d}^4k_1}{(2\pi)^{4}}(2\pi)\delta^+(k_1^2-m^2)S^\dagger (\sigma(k_1))\langle q_n|T^{\dagger}|k_1\rangle 
\int\frac{\mathrm{d}^4k_2}{(2\pi)^{4}}(2\pi)\delta^+(k_2^2-m^2)S(\sigma(k_2))\langle k_2|T|p_m\rangle \\\nonumber
&\times v(k_1,P-k_1-k_2)
(2\pi)\delta^+((P-k_1-k_2)^2-m^2)
v(P-k_1-k_2,k_2)\,. \hspace{15cm}
\end{align}
The corresponding topology on the right panel of Fig.~\ref{fig:compare} reads
\begin{align}
\label{eq:bse4b}
\langle q_1,q_2,q_3|&(\hat T-\hat T^{\dagger})| p_1,p_2,p_3\rangle_{{\rm (4b)}}
\\\nonumber
=&i\frac{1}{3!}\sum_{n,m}\int\frac{\mathrm{d}^4k_1}{(2\pi)^{4}}\int\frac{\mathrm{d}^4k_2}{(2\pi)^{4}}v(q_{\bar{n}},q_{\barr{n}})S^\dagger (\sigma(q_n))\langle q_n|T^{\dagger}|k_1\rangle \tau^{\dagger}(\sigma(k_1)) \\\nonumber 
&\times\left(\langle k_1|B|k_2 \rangle - \langle k_1|B^\dagger|k_2 \rangle\right) \tau (\sigma(k_2)) \langle k_2|T|p_m\rangle S(\sigma(p_m))v(p_{\bar{m}},p_{\barr{m}}) \\\nonumber
=&i\frac{1}{3!}\sum_{n,m}v(q_{\bar{n}},q_{\barr{n}})S^\dagger (\sigma(q_n))S(\sigma(p_m))v(p_{\bar{m}},p_{\barr{m}}) 
\\\nonumber
&
\times\int\frac{\mathrm{d}^4k_1}{(2\pi)^{4}}\tau^{\dagger}(\sigma(k_1)) \langle q_n|T^{\dagger}|k_1\rangle\int\frac{\mathrm{d}^4k_2}{(2\pi)^{4}}\tau (\sigma(k_2)) \langle k_2|T|p_m\rangle 
\times (\langle k_1|B|k_2 \rangle - \langle k_1|B^\dagger|k_2 \rangle)\,, \hspace{15cm}
\end{align}
The latter both equations hold for arbitrary choice of external momenta. Therefore, element-wise comparison using again the assumption ${\tau(\sigma(k)) = (2\pi)\delta^+(k^2-m^2)S(\sigma(k))}$ yields
\begin{align}
\label{eq:result4b}
\langle k_1|B|k_2 \rangle - \langle k_1|B^\dagger|k_2 \rangle=
iv(P-k_1-k_2,k_1)(2\pi)\delta^+((P-k_1-k_2)^2-m^2)v(P-k_1-k_2,k_2)\,.
\end{align}
All three expressions of the discontinuity of the driving term of the Bethe-Salpeter equation are consistent among each other with the only difference being the values that the spectator momentum can take. The most general expression is, therefore, given by Eq.~\eqref{eq:result4b}, which holds for arbitrary momenta constrained only by $k^2=m^2$.

\section{Discontinuity of the isobar-spectator scattering amplitude}
\label{sec:appendix_b}

The Bethe-Salpeter equation has two equivalent expressions, which in operator language read
\begin{align}
T = B + B\tau T \text{~~~or~~~} B = T(1 + \tau T)^{-1}\,.
\end{align}
Writing the discontinuity of $B$ and multiplying it from the left with 
$(1 + T^\dagger\tau^\dagger)$ and from the right with $(1 + \tau T)$ one obtains
\begin{align*}
T(1 + \tau T)^{-1}-(1 + T^\dagger \tau^\dagger)^{-1}T^\dagger&=B-B^\dagger
\\\nonumber
(1 + T^\dagger\tau^\dagger)T - T^{\dagger}(1 + \tau  T)
&=
(1 + T^\dagger\tau^\dagger)(B - B^\dagger)(1 + \tau  T)  
\\\nonumber
T-T^{\dagger} - T^{\dagger}(\tau  - \tau^{\dagger})T
&=
(B - B^\dagger) + T^{\dagger}\tau^{\dagger}(B - B^\dagger) + (B - B^\dagger)T\tau  + T^{\dagger}\tau^{\dagger}(B - B^\dagger)T\tau \,.
\end{align*}
This produces immediately Eq.~(\ref{eq:bse3}).


\section{Limit of a narrow resonance}
\label{app:limit}
As discussed at the end of Sec.~\ref{sec:final} one can add a real contribution $V$ to the interaction $B$ without spoiling unitarity. With such a $\lambda$-independent $V$ the amplitude $T$ remains finite in the limit $\lambda\to 0$ and the transition of a resonance isobar becoming a stable particle can be studied. Then, Eq.~(\ref{eq:final}) reads
\begin{align}\label{eq:narrow}
\langle q|T(s)|p\rangle =
& 
\langle q|V|p\rangle
+
\int
\frac{\mathrm{d}^3\boldsymbol{\ell}}{(2\pi)^3}\langle q|V|\ell\rangle\,G(\boldsymbol{\ell})\,\langle\ell|T(s)|p\rangle\ ,\quad 
G(\boldsymbol{\ell}):=\frac{1}{2E_\ell(2\sqrt{s}\,E_\ell-s-m^2+M_0^2-i\epsilon)} \, .
\end{align}
The imaginary part is given by
\begin{align}
{\rm Im}\,G(\boldsymbol{\ell})=\frac{\pi\delta(|\boldsymbol{\ell}|-\ell_{\rm cm})}{4\ell_{\rm cm}\sqrt{s}} \ ,
\end{align}
where $\ell_{\rm cm}$ is the center-of-mass momentum of the particles with masses $m$ and $M_0$. The same imaginary part results from the two-particle Bethe-Salpeter equation which in the current convention ($\mathcal{S}{}:=\mathbbm{1}+i(2\pi)^4 \delta^4\!\left(P_{i}-P_{f}\right)  T$) reads~\cite{Aaron:1969my}
\begin{align}
\label{eq:tpBSE}
\langle q|T(s)|p\rangle =
& 
\langle q|V|p\rangle
+
\int
\frac{\mathrm{d}^3\boldsymbol{\ell}}{(2\pi)^3}\frac{\langle q|V|\ell\rangle}{4E_\ell\omega_\ell}\left(\frac{1}{E_\ell+\omega_\ell-\sqrt{s}-i\epsilon}+\frac{1}{\sqrt{s}+E_\ell+\omega_\ell}\right)\langle\ell|T(s)|p\rangle \, 
\end{align}
after integration of $\ell_0$. Here, $\omega_\ell^2=M_0^2+|\boldsymbol{\ell}|^2$. 
Thus, we indeed obtain a two-particle unitary scattering equation in the limit $\lambda\to 0$ as expected.
A closer inspection shows that the re-scattering term in Eq.~(\ref{eq:narrow}) corresponds to the first term in Eq.~(\ref{eq:tpBSE}), while the second term that does not become singular has no counterpart in Eq.~(\ref{eq:narrow}).

\bibliography{quattro}

\end{document}